# Interaction of Luminescent Defects in Carbon Nanotubes with Covalently Attached Stable Organic Radicals

*Felix J. Berger[1,2], J. Alejandro de Sousa[3,4], Shen Zhao[5,6], Nicolas F. Zorn[1,2], Abdurrahman Ali El Yumin[1,2], Aleix Quintana García[3], Simon Settele[1], Alexander Högele[5,6], Núria Crivillers[3] and Jana Zaumseil[1,2]\**

[1]Institute for Physical Chemistry, Universität Heidelberg, 69120 Heidelberg, Germany

[2]Centre for Advanced Materials, Universität Heidelberg, 69120 Heidelberg, Germany

[3]Institut de Ciència de Materials de Barcelona (ICMAB-CSIC), Campus UAB, 08193 Bellaterra, Spain

[4]Laboratorio de Electroquímica, Departamento de Química, Facultad de Ciencias, Universidad de los Andes, 5101 Mérida, Venezuela

[5]Faculty of Physics, Munich Quantum Center and Center for NanoScience (CeNS), Ludwig-Maximilians-Universität München, 80539 München, Germany

[6]Munich Center for Quantum Science and Technology (MCQST), 80799 München, Germany

Corresponding Author

\* E-mail: zaumseil@uni-heidelberg.de

**ABSTRACT**

The functionalization of single-walled carbon nanotubes (SWCNTs) with luminescent $sp^3$ defects has greatly improved their performance in applications such as quantum light sources and bioimaging. Here, we report the covalent functionalization of purified semiconducting SWCNTs with stable organic radicals (perchlorotriphenylmethyl, PTM) carrying a net spin. This model system allows us to use the near-infrared photoluminescence arising from the defect-localized exciton as a highly sensitive probe for the short-range interaction between the PTM radical and the SWCNT. Our results point toward an increased triplet exciton population due to radical-enhanced intersystem crossing, which could provide access to the elusive triplet manifold in SWCNTs. Furthermore, this simple synthetic route to spin-labeled defects could enable magnetic resonance studies complementary to *in vivo* fluorescence imaging with functionalized SWCNTs and facilitate the scalable fabrication of spintronic devices with magnetically switchable charge transport.

The controlled covalent modification of single-walled carbon nanotubes (SWCNTs) has boosted their potential for a broad range of applications including quantum light emission,[1] bioimaging[2] and sensing.[3] The one-dimensional structure of SWCNTs, their large charge carrier mobilities[4] and high exciton diffusivity[5] make them ideal for efficient interaction of charges and excitations with lattice defects, thus further expanding the nanotubes' properties. In recent years, the low-density functionalization of semiconducting SWCNTs with luminescent sp$^3$ defects, sometimes referred to as organic color centers or quantum defects, has received particular attention.[6-8] The deep optical trap potentials (~100 meV) of these defects created by binding of oxygen,[9] aryl[10] or alkyl[11] substituents lead to efficient exciton localization. These trapped excitons give rise to red-shifted photoluminescence (PL) with long lifetimes and even enable room-temperature single-photon emission at telecommunication wavelengths.[12] As the defect states arise from the perturbation of the SWCNTs' electronic structure, their properties are dominated by the precise binding pattern with respect to the SWCNT lattice rather than the molecular structure of the substituents.[13] Hence, the molecular groups that are bound to the nanotube surface are often structurally simple and only few examples are known, in which the attached group contributes and performs a function itself, such as sensing[14-16] or anchoring of biomolecules.[17]

A highly desirable property for functional groups attached to SWCNTs is a net spin, as carried by unpaired electrons. Previously, mixed-chirality samples of SWCNTs and multi-walled carbon nanotubes were decorated with open-shell transition metal[18] or lanthanide[19, 20] complexes and charge transport measurements at cryogenic temperatures revealed spin valve switching behavior without the need for ferromagnetic contacts.[19, 20] Another class of spin-bearing compounds are stable organic radicals, *i.e.*, charge-neutral molecules with an unpaired electron.[21, 22] One of the most prominent examples is the perchlorinated triphenylmethyl (PTM) radical, in which a propeller-shaped arrangement of phenyl rings protects the carbon-centered radical resulting in a half-life of ~100 years.[23] The interaction of SWCNTs with the unpaired spin of the radical *via* a luminescent sp³ defect may result in spin-dependent properties and provide optical means for probing them. However, up to now there are no reports of stable organic radicals bound to carbon nanotubes or chirality-sorted SWCNTs with any covalent spin label.

In this work, we functionalize purely semiconducting (6,5) SWCNTs with aryl defects linked to PTM radicals and use the PL arising from the defect-localized exciton as a highly sensitive probe for the short-range interaction between the radical and the SWCNT. By comparing them to aryl defects with closed-shell substituents, we observe PL quenching associated with the radical. Based on the PL decay dynamics, we propose that partial transfer of the exciton population to triplet states *via* radical-enhanced intersystem crossing is the dominant quenching mechanism, while a smaller portion may be attributed to a photo-induced electron transfer process. In addition to providing a simple route to spin-labeled defects for magnetic resonance studies and spintronics, radical-functionalization could improve the accessibility of the elusive triplet states in carbon nanotubes.

**RESULTS AND DISCUSSION**

**SWCNT functionalization and characterization**

Highly purified (6,5) SWCNTs (diameter ~0.76 nm, average length 1.0 - 1.5 µm) wrapped with the fluorene-bipyridine copolymer PFO-BPy (see the **Methods** and **Figure 1** for molecular structure)[24] serve as a robust model system for sp³ functionalization. As shown in **Figure 1**, diazonium (Dz) chemistry was used to introduce aryl defects bearing charge-neutral substituents with either radical (open-shell) or closed-shell character. Reacting (6,5) SWCNTs with the tailored diazonium salt PTM-Dz[25, 26] produced aryl defects on the nanotube lattice that act as covalent linkers to PTM radicals. To identify the impact of the radical on defect state PL, the closed-shell hydrogen-substituted PTMH-Dz was employed to create reference defects. Additionally, simple closed-shell bromoaryl defects (ArBr) were produced by reaction with the commercial Br-Dz reagent, which exhibited a higher reactivity than the PTMH-Dz and was therefore used for experiments requiring larger defect densities. Since polymer-wrapped SWCNTs form stable dispersions only in low-polarity solvents, the functionalization with Br-Dz was facilitated by ether crown complexation in a toluene/acetonitrile mixture following our recently reported protocol.[27] In contrast to that, PTM-Dz and PTMH-Dz are well-soluble in tetrahydrofuran (THF), which also disperses PFO-BPy-wrapped SWCNTs. They are therefore ideally suited for the reaction with polymer-wrapped SWCNTs without the need for any solubilizing agents. Nanotube functionalization was always performed in the dark and completed within 16 hours at room temperature. Subsequently, the functionalized SWCNTs were collected by vacuum filtration, carefully washed with pure solvent to remove unreacted diazonium salt and byproducts and finally redispersed in fresh toluene for characterization.

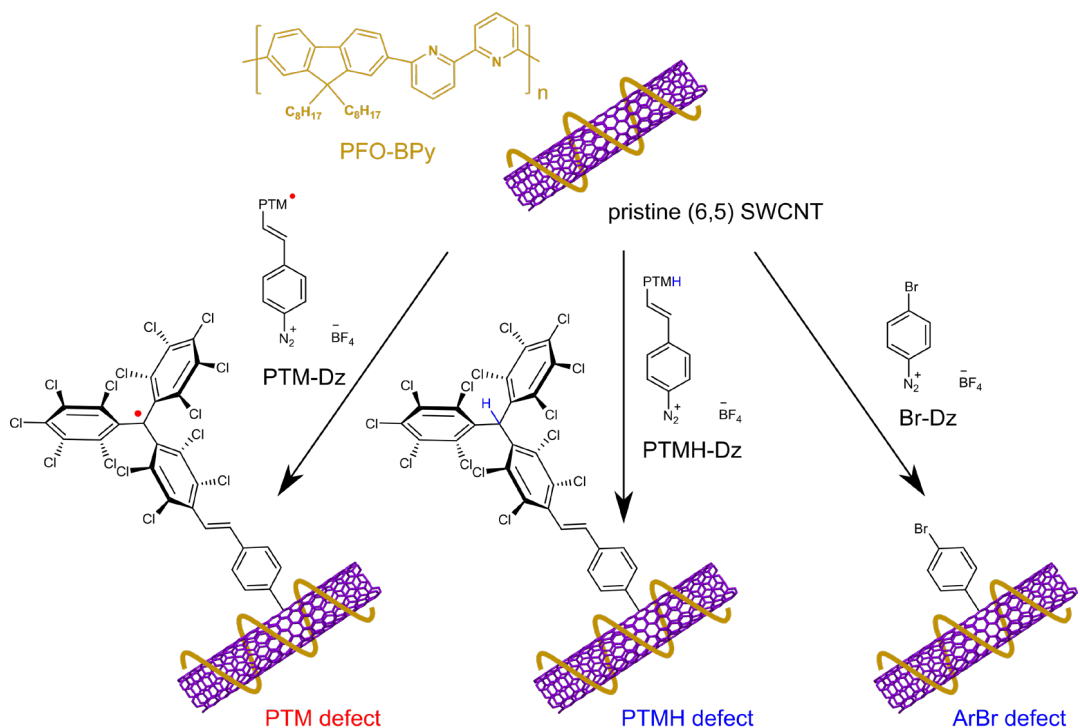

**Figure 1.** Reaction scheme depicting the functionalization of polymer-wrapped (6,5) SWCNTs with aryl defects bearing radical (PTM, red) or closed-shell (PTMH, ArBr, blue) substituents.

To confirm the covalent functionalization of (6,5) SWCNTs with PTM radicals, photoluminescence and electron paramagnetic resonance (EPR) spectra were recorded for toluene dispersions of the nanotubes at room temperature (**Figure 2**). Here, we focus on a comparison between PTM radical-functionalized SWCNTs and the corresponding closed-shell PTMH-functionalized SWCNTs as well as pristine SWCNTs. All (6,5) SWCNT dispersions exhibit photoluminescence at 1006 nm originating from mobile excitons recombining on pristine lattice segments ($E_{11}$ transition). In addition, both sp$^3$ functionalized nanotubes show emission at ~1165 nm, which is characteristic of defect-localized excitons ($E_{11}$*) and strong evidence for the covalent attachment of molecular groups to the nanotube lattice (**Figure 2a**). To analyze the influence of the radical substituent on the defect state energy, the optical trap depth[28] ($E_{11}$-$E_{11}$*) was extracted from the PL spectra of PTM- and PTMH-functionalized

SWCNTs (**Supporting Information Figure S1**). The optical trap depth of PTM defects (172 meV) was slightly larger than that of PTMH defects (168 meV), which is in agreement with the stronger electron-withdrawing character of the PTM group.[10] However, the difference is small, presumably due to the relatively long distance between the central carbon of the PTM moiety and the SWCNT lattice. We can conclude that the optical trap depth should only play a minor role with regard to any differences observed between PTM and PTMH defects.

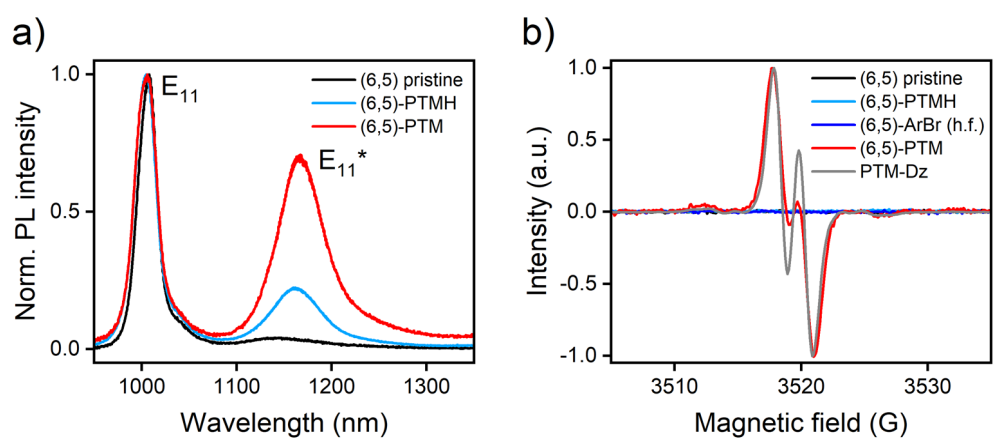

**Figure 2.** (a) Normalized PL spectra recorded for dispersions of pristine, PTM-, PTMH-functionalized (6,5) SWCNTs in toluene. (b) EPR spectra (at room temperature) of these dispersions in addition to a dispersion of (6,5) SWCNTs with a high density of ArBr defects (h.f.) and a solution of the PTM-Dz precursor in tetrahydrofuran.

While the presence of an $E_{11}^*$ emission feature clearly indicates covalent functionalization, it is relatively insensitive to the precise molecular structure of the attached group.[13] Hence, EPR measurements were performed to ascertain that the substituent was indeed the desired PTM-styryl unit and the radical character was preserved after the reaction. **Figure 2b** shows that the pristine (6,5) SWCNTs do not display any EPR signal, as expected for highly purified semiconducting SWCNTs.[29] Likewise, closed-shell PTMH-functionalized SWCNTs show no EPR intensity at room temperature. To make sure that this result was not compromised by the low density of PTMH defects, a second control experiment was performed on (6,5) SWCNTs that were highly functionalized with closed-shell ArBr defects, which also did not give rise to any EPR signal (see **Supporting Information Figure S2** for additional characterization). The PTM-functionalized SWCNTs, on the other hand, show the clear EPR signature of the PTM-styryl unit with a g factor of 2.0024, which is characteristic for PTM radicals, and a peak splitting due to hyperfine coupling between the PTM radical and the vinylene proton with appropriate symmetry (for more details on the fine structure see the **Supporting Information Figure S3**). Furthermore, the EPR signal of the PTM radical attached to the SWCNT is broadened compared to that of the molecular precursor PTM-Dz (linewidths of 1.4 G and 1.1 G, respectively, see **Supporting Information Figure S4**). This broadening may be attributed to the slow tumbling[30] of the micrometer-long nanotube in dispersion with a rotational correlation time of ~10 ms rad$^{-1}$,[31] which is much longer than that of the unbound PTM-Dz molecule. Consequently, the line broadening further corroborates the successful decoration of SWCNTs with stable organic radicals.

These results clearly indicate that the room temperature EPR response of PTM-tailored SWCNTs originates from an unpaired spin localized on the PTM moiety and not on the nanotube itself. In contrast to that, Lohmann *et al.*[32] recently reported that aryl-functionalization of SWCNTs with closed-shell substituents in aqueous dispersions may give rise to an EPR signal at low temperatures. The authors assigned this signal to a nanotube-centered radical, which might be formed upon addition of an odd number of sp$^3$ defects to the lattice. Since the intensity of this EPR signal decreased strongly with increasing temperature and became undetectable above 200 K,[32] such a feature is not expected to appear in our room temperature EPR measurements.

**Impact of radical on defect PL**

The unusual observation of a pronounced absorption band accompanied by relatively weak PL from the PTM defects (compare **Figure 2** and **Supporting Information Figure S2**) prompted a more detailed investigation of their PL efficiency in relation to a reference sample of closed-shell ArBr-functionalized SWCNTs with similar defect density as indicated by the $E_{11}^*/E_{11}$ absorbance ratio (**Figure 3a**). As shown in **Figure 3b**, the radical-functionalized defects exhibit a significantly lower $E_{11}^*/E_{11}$ PL intensity ratio than the closed-shell reference despite having a similar defect density. This picture is complemented by PL lifetime measurements *via* time-correlated single-photon counting (TCSPC) for the $E_{11}^*$ PL from different defect types. The data indicate a substantially shorter decay time for the radical-functionalized defects compared to both closed-shell references (**Figure 3c**). It should be noted that the emission from all samples was filtered by a grating spectrograph to pass only the intensity from a narrow wavelength band around 1165 nm to the single-photon detector. Thereby, we rule out the wavelength-dependence of defect state lifetime as a possible cause for the observed differences.[33] Combined with the reduced defect emission intensity, the shortened PL lifetime of PTM-functionalized SWCNTs points toward a non-radiative relaxation pathway associated with the radical in close proximity to the defect. Notably, the $E_{11}^*$ PL decay of the radical-functionalized SWCNTs can still be described by the biexponential model typical for localized exciton decay dynamics at room temperature.[33, 34]

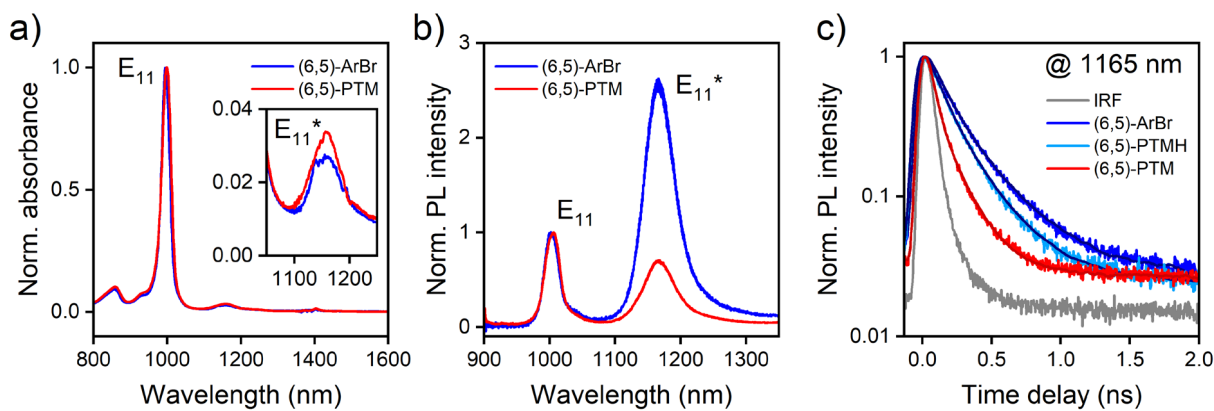

**Figure 3.** (a) Absorption and (b) PL spectra of PTM- and ArBr-functionalized (6,5) SWCNTs with similar defect density. (c) PL decay traces of defect emission at 1165 nm. All data were measured on SWCNT dispersions in toluene.

If the observed $E_{11}^*$ PL quenching results from the presence of the radical substituent on the aryl defect, the *in situ* conversion of open-shell to closed-shell substituents on the sp$^3$ defects will be the most direct way to confirm and quantify this effect. For this purpose, we exploited the fact that PTM radicals are sensitive to UV and blue light (in the range of their absorption maximum around 385 nm). It is well-established that irradiation in this spectral range induces chlorine atom elimination from the PTM followed by ring closure to the perchlorophenylfluorenyl (PPF) radical.[35, 36] Due to the planar structure of the fluorene unit hosting the radical, the PPF radical is much more susceptible to chemical attack than the PTM radical. Hence, we consider hydrogen atom abstraction from the solvent and formation of the closed-shell PPFH species as a plausible decomposition pathway, as shown schematically in **Supporting Information Figure S5**. In practice, degradation of the PPF radical yields a mixture of closed-shell products. Although the extremely low defect concentrations (estimated to be in the range of nmol/L) did not allow us to determine the molecular structure of any photolysis products, we confirmed experimentally that all of them were of closed-shell nature (see below). As noted in the discussion of optical trap depths of PTM *versus* PTMH defects, minor changes in substituent molecular structure far away from the SWCNT do not impact the $E_{11}^*$ PL properties significantly,[13] thus the PPFH substituent is a good closed-shell reference system for the open-shell PTM.

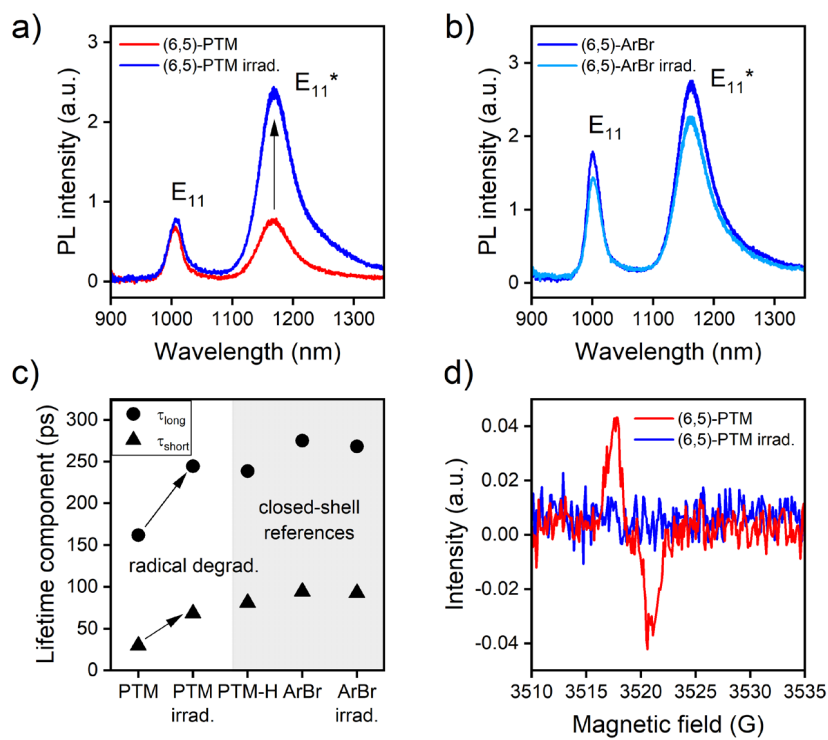

**Figure 4.** (a) PL spectra of PTM-functionalized and (b) ArBr-functionalized (6,5) SWCNTs before and after irradiation. (c) PL lifetime components extracted from biexponential fits to the decay traces. (d) EPR spectra recorded for the dispersions used in the PL experiments before and after irradiation. Note that the initial EPR signal intensity is relatively low, because dilute dispersions were employed for the optical characterization.

For *in situ* conversion, a dispersion of PTM-functionalized SWCNTs was exposed to UV light (365 nm, 0.2 W cm$^{-2}$) for 40 min and PL spectra as well as PL decay traces were recorded under identical conditions before and after irradiation. As expected, the $E_{11}^*$ PL intensity strongly increased due to light-induced radical degradation, whereas the $E_{11}$ intensity remained constant (**Figure 4a**, note that absolute PL intensities are compared). Hence, the creation of additional $sp^3$ defects as the origin of increased $E_{11}^*$ intensity can be ruled out. In that case, the $E_{11}$ intensity would be reduced due to the higher density of exciton trapping sites. The unchanged $E_{11}$ emission confirms that the interaction between the radical and the SWCNT is confined to the $sp^3$ defect. To exclude that these changes are related to photochemical reactions on the SWCNT lattice itself, *e.g.*, re-arrangement of $sp^3$ defects, the experiment was repeated with a dispersion of ArBr-functionalized SWCNTs. As shown in **Figure 4b**, these defects did not respond to UV irradiation. The minor drop of PL intensity is likely due to slight SWCNT aggregation over the course of irradiation and the normalized spectra in **Supporting Information Figure S6** illustrate that the PL spectral shape does not change at all. Hence, the $E_{11}^*$ brightening observed in **Figure 4a** can be attributed unambiguously to the photochemical transformation of the PTM radical to a closed-shell system.

The marked increase in $E_{11}^*$ emission efficiency can be understood by considering the prolonged PL lifetime of the defect-localized exciton following the open-shell to closed-shell conversion (**Figure 4c**). Upon irradiation, the initially short PL lifetime of the PTM defects rises to the level of the closed-shell references (PTMH and ArBr). By comparing EPR spectra of the PTM-functionalized SWCNT dispersion before and after irradiation (**Figure 4d**), we can conclude that the PTM radical was indeed converted to a closed-shell species. Consequently, the mechanism responsible for $E_{11}^*$ PL quenching in PTM-functionalized SWCNTs must be directly linked to the presence of the radical on the substituent of the aryl defect. The PL quenching effect can be quantified based on the PL spectral and lifetime data. **Table 1** details the extracted lifetime components and amplitudes together with the integrated $E_{11}^*$ to $E_{11}$ PL intensities for each defect type. We find that the amplitude-averaged $E_{11}^*$ lifetime ($\tau_{amp-av}$) of PTM defects increases by a factor of 2.4 upon radical decomposition, which is in excellent agreement with the 2.5-fold enhancement of the $E_{11}^*/E_{11}$ PL intensity ratio.

**Table 1.** PL lifetime components (long: $\tau_l$, short: $\tau_s$) with corresponding amplitudes ($A_l$ and $A_s$), amplitude-averaged lifetimes ($\tau_{amp-av}$) and integrated defect-to-$E_{11}$ intensity ratio PL($E_{11}^*/E_{11}$).

| Defect | $A_l$ (%) | $\tau_l$ (ps) | $A_s$ (%) | $\tau_s$ (ps) | $\tau_{amp-av}$ (ps) | PL($E_{11}^*/E_{11}$) |
|---|---|---|---|---|---|---|
| PTM | 16 | 162 | 84 | 30 | 51 | 2.6 |
| PTM irrad. | 30 | 245 | 70 | 68 | 121 | 6.5 |
| PTMH | 29 | 239 | 71 | 81 | 127 | 0.6 |
| ArBr | 31 | 275 | 69 | 95 | 150 | 3.7 |

**PL quenching mechanism**

In the following, we discuss the possible underlying mechanisms of quenching of the defect-localized exciton by the radical substituent on the aryl defect. Three mechanisms must be considered: Resonant excitation energy transfer, photo-induced electron transfer and enhanced intersystem crossing. First, the possibility of energy transfer from the sp$^3$ defect state to the PTM radical can be ruled out based on the much smaller optical gap of the defect compared to the PTM and negligible spectral overlap between the two partners as shown in **Supporting Information Figure S7**.

Second, in order to assess the potential for photo-induced electron transfer (PET), the redox potentials of all involved species and their optical transition energies must be considered (see **Table 2**). From this data, the Gibbs free energy for PET from pristine or sp$^3$-functionalized regions of the SWCNT (donor D) to the PTM (acceptor A) was evaluated according to[37]

$$\Delta G_{PET} = eE_{red}(D^+/D) - eE_{red}(A/A^-) - \Delta G_{00} - \frac{e^2}{4\pi\varepsilon_r\varepsilon_0 d}$$

where $E_{red}$ denotes the reduction potential, $\Delta G_{00}$ is the optical transition energy and the electrostatic work term amounts to 0.5 eV for the relative permittivity of toluene ($\varepsilon_r = 2.38$) and an ion pair separation $d \approx 1.2$ nm. Following PET to the functional group, the electron is back-transferred to the ground state of the defect/SWCNT, thereby resulting in non-radiative recombination. As indicated by a large negative $\Delta G_{PET}$ in **Table 2**, PET is thermodynamically highly favorable[38] and could potentially quench both $E_{11}$ and $E_{11}$* emission. However, since the rate of electron transfer reactions decreases exponentially on a length scale of ~0.1 nm,[39] the quenching of mobile $E_{11}$ excitons can be excluded due to their long average distance to the radical, in agreement with negligible $E_{11}$ brightening in **Figure 4a**. The defect-localized exciton, on the other hand, is sufficiently close to the PTM to be quenched *via* PET, even though the separation of ~1 nm is likely to limit the rate of this process. For the sake of completeness, it is worth noting that the redox potentials in **Table 2** permit ground state charge transfer from the SWCNT/defect to the PTM as well, but such a permanent ion pair formation at the defect site is irreconcilable with the observation of the EPR signature from the PTM defect and the PL lifetime shortening, which indicates dynamic rather than static quenching.

**Table 2.** Redox potentials ($E_{ox}$, $E_{red}$), optical transition energies ($\Delta G_{00}$) and Gibbs free energies for photo-induced electron transfer ($\Delta G_{PET}$).

| Species | $E_{ox}$ (V)[a] | $E_{red}$ (V)[a] | $\Delta G_{00}$ (eV)[c] | $\Delta G_{PET}$ (eV)[d] |
|---|---|---|---|---|
| (6,5) pristine | 0.615 | -0.420 | 1.24 | -2.17[e] |
| (6,5) sp³ defect[b] | 0.584 | -0.398 | 1.06 | -1.95[f] |
| PTM | 1.61 | -0.19 | - | - |

[a]Potentials from Shiraishi et al.[40] and Souto et al.[41] measured *versus* a Ag/AgCl reference electrode.
[b]Data from Shiraishi et al.[40] for ArBr defects was used as an approximation to PTM-substituted aryl defects because of their near-identical optical trap depths.
[c]Energy of the $E_{11}$ and $E_{11}^*$ PL transition, respectively.
[d]Including an electrostatic work term of -0.5 eV.
[e]For the reaction: SWCNT + PTM → SWCNT$^+$ + PTM$^-$.
[f]For the reaction: defect + PTM → defect$^+$ + PTM$^-$.

As another potential mechanism, covalently bound radicals have been shown to substantially speed up intersystem crossing in organic fluorophores.[42-45] This effect can be modeled by treating the electronic exchange interaction as a perturbation that promotes spin flips of the unpaired electron on the radical and the photo-excited electron on the chromophore.[46] The intersystem crossing rate thus depends sensitively on the degree of wavefunction overlap between the singly-occupied molecular orbital (SOMO) on the radical and the singlet exciton. It can reach timescales of ~1-100 ps for not too distant radicals.[42-45] Hence, this mechanism is similarly short-ranged as PET and could affect defect-localized excitons close to the PTM radical. As discussed below, such radical-enhanced intersystem crossing (EISC) would allow triplet states to participate in the fast population redistribution process between bright and dark defect states following exciton trapping.[33, 34, 47] In this case, the radical-induced PL quenching may be explained by exciton population transfer from the bright singlet to a dark triplet state.

Both the PET and the EISC mechanism are likely to occur for PTM defects based on thermodynamic considerations and previous work on structurally similar molecular systems. To further corroborate or exclude the involvement of either quenching mechanism, we now discuss the $E_{11}$* PL decay in more detail as it is rich in information and data are available for both radical-functionalized and reference defects (**Table 1**). Hartmann *et al.*[34] and He *et al.*[33] developed a model for the biexponential decay of excitons localized at defects with closed-shell substituents at room temperature. Since the $E_{11}$* PL decay of radical-functionalized defects remains biexponential with well-separated long and short timescales, we assume that the dynamics in such defects are still similar to the existing model but should be modified by introducing PET or EISC. With these assumptions, the observed radical-induced changes in the long and short decay components can only be explained by a combination of PET and reversible EISC to a triplet state that is energetically below the bright state, as discussed in detail in the **Supporting Information**.

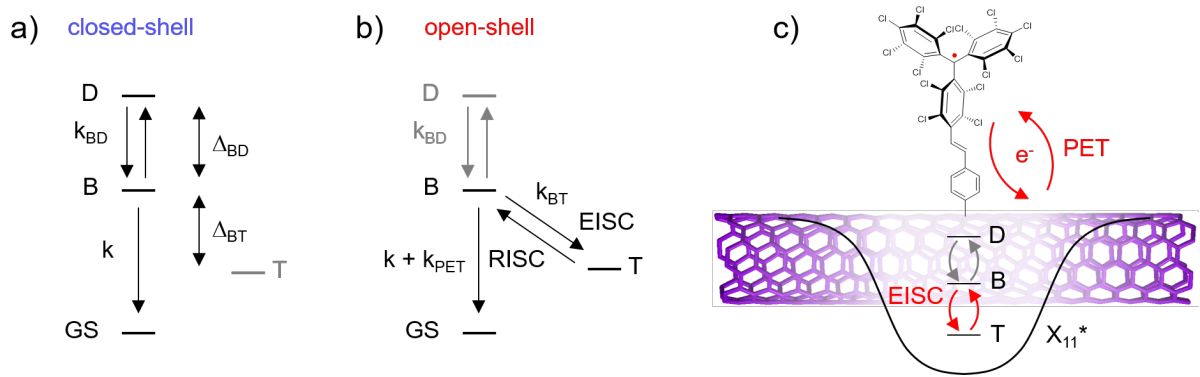

**Figure 5.** Model of exciton dynamics in SWCNTs with sp³ defects with (a) closed-shell or (b) open-shell functional groups. The energy level diagrams include the ground state (GS), the defect-localized bright (B) and dark (D) singlet states and a triplet (T) state with energy gaps $\Delta_{BD}$ and $\Delta_{BT}$. The relevant rate constants are indicated as follows: k - radiative and non-radiative recombination; $k_{PET}$ - photo-induced electron transfer (PET); $k_{BD}$ - equilibration between B and D state population; $k_{BT}$ - equilibration between B and T state population *via* radical-enhanced intersystem crossing (EISC) and thermally activated reverse intersystem crossing (RISC). Note that $k_{BD}$ and $k_{BT}$ are not the transition rates between states, but represent the relaxation rates at which the involved populations approach their equilibrium value. The rate of this relaxation process depends on the elementary transition rates. Gray color marks states with small population or rates that are not competitive. (c) Schematic illustration of PET and EISC as the PL quenching processes in open-shell PTM defects.

**Figure 5** illustrates our proposed model for defects with open-shell substituents and compares it with the model that is generally applied for defects with closed-shell substituents.[33, 34, 47] The involved states are the defect-localized bright singlet exciton (B) and parity-forbidden dark singlet exciton (D), as well as a triplet exciton (T), which could be either mobile or defect-localized. These states are assumed to reversibly exchange population at room temperature as the energy gaps between them are either smaller ($\Delta_{BD} \approx 9$ meV as found by Kim *et al.*[47]) or not much larger than 25 meV (an upper boundary of $\Delta_{BT} \lesssim 38$ meV will be estimated below). In general, the fast decay component is attributed to the redistribution of exciton population among these bright and dark states, which is followed by the slow recombination of trapped excitons *via* radiative and non-radiative channels. More specifically, as intersystem crossing is slow in the absence of the radical (**Figure 5a**), only the D state accepts population from the B state with the redistribution rate $k_{BD}$. Subsequently, trapped excitons recombine *via* multiphonon decay and other pathways with the cumulative rate $k = \tau_l^{-1}$ while the ratio of B and D state populations retains its equilibrium value. Since recombination can occur at any time after the trapping event, the short decay component is given by $\tau_s^{-1} = k_{BD} + \tau_l^{-1}$. In contrast to that, we propose that radical-functionalization (**Figure 5b**) enables fast EISC and thermally activated reverse intersystem crossing (RISC) between the B state and an energetically low-lying T state. As a result, this T state dominates the population redistribution process among bright and dark states (for details see **Supporting Information**). Simultaneously, PET is responsible for enhanced non-radiative recombination, resulting in $\tau_l^{-1} = k + k_{PET}$ and $\tau_s^{-1} = k_{BT} + \tau_l^{-1}$ for radical-substituted defects.

A quantitative analysis of the data in **Table 1** further supports this picture. Assuming that PET is responsible for changes in the decay rates $\tau_l^{-1}$, a PET rate of $\tau_l^{-1}$(PTM) - $\tau_l^{-1}$(PTM irrad.) = $(500 \text{ ps})^{-1}$ is extracted. This moderately fast electron transfer rate is rationalized by the trade-off between the large thermodynamic driving force for the process ($\Delta G_{PET}$ = -1.95 eV) and the relatively long distance (~1.2 nm) between the SWCNT surface and the radical center. Further, the term ($\tau_s^{-1} - \tau_l^{-1}$) yields the population redistribution rates $k_{BD} = (91 \text{ ps})^{-1}$ and $k_{BT} = (37 \text{ ps})^{-1}$ for the closed- and open-shell case, respectively. Note that the calculated $k_{BT}$ is in good agreement with EISC rates[43, 45] for chromophore-radical distances of 1 nm as present in PTM defects. As pointed out by He *et al.*,[33] the normalized amplitude of the long decay component may be used as a measure of bright state population in equilibrium with respect to its initial population immediately after trapping (**Supporting Information**). For radical-functionalized defects $A_l$ is 16 %, whereas $A_l$ is about 30 % across all reference defects in good agreement with previous reports.[33] This reduction of $A_l$ further supports EISC-mediated population transfer from the B to the T state as a contribution to $E_{11}$* PL quenching. Moreover, because the equilibrium populations of the B, D and T state should simply reflect the Boltzmann distribution, we can use the information that $A_l$ = 16 % together with $\Delta_{BD} \approx 9$ meV for $E_{11}$* defects in (6,5) SWCNTs[47] to obtain a rough estimate of $\Delta_{BT} \approx 38$ meV. So far, we assumed a single triplet state to be involved, although several triplet states are predicted[48] within a few 10 meV of the bright $E_{11}$* singlet. Including a second triplet state will further reduce $\Delta_{BT}$, so 38 meV should be regarded as an upper boundary, which justifies the earlier assumption of RISC at room temperature. Finally, we note that within this model about ~1/3 of the defect PL intensity drop results from PET, whereas ~2/3 are a consequence of EISC (**Supporting Information**). It is important to emphasize that our description of the decay dynamics of radical-functionalized defects and the extracted physical parameters only apply as long as the underlying model for closed-shell defects[33, 34] is valid. For example, if the biexponential decay instead resulted from shelving into dark states, similar to the case of $E_{11}$ excitons,[49, 50] other quenching mechanisms might match the data more closely.

As long lifetimes (> 1 µs) are a hallmark of triplet states in organic materials, we also looked for delayed fluorescence at room temperature.[51, 52] To this end, the PL decay of PTM-functionalized SWCNTs was monitored at the $E_{11}$ or $E_{11}$* peak emission wavelength by TCSPC over a time window of 0.3 µs. However, in neither case a component with µs lifetime was detectable. While it is possible that the delayed fluorescence signal is simply too weak to be detected in our setup, it is important to keep in mind that RISC (from the T to the B state) is enhanced by the same factor as the forward process. Thus, their efficient coupling *via* EISC and RISC should lead to simultaneous depopulation of the B and the T state on the timescale of $\tau_l$ (~100 ps). Hence, the effective triplet state lifetime is strongly reduced and delayed fluorescence is no longer expected.

**Magneto-photoluminescence spectroscopy**

Given the clear evidence for interaction of the radical with the trapped exciton, we explored how the paramagnetic PTM substituent influences the PL of sp$^3$ defects in magnetic fields. Low-temperature magneto-PL experiments were performed on individual, PTM-functionalized (6,5) SWCNTs embedded in a polystyrene layer (see **Supporting Information Figure S8** for a scanning confocal PL image). As illustrated in **Figure 6a**, the sample was mounted such that PL could be collected while applying an in-plane magnetic field (Voigt geometry). The random nanotube orientation results in a tube-specific angle between the tube axis and the magnetic field. In total, eight nanotubes were investigated that displayed single-line or well-separated multi-line defect PL between 1120 and 1180 nm (1.05-1.11 eV). Note that the sidebands that are red-shifted by 2-3 meV from the zero-phonon line are typical of the cryogenic PL spectra of polymer-wrapped SWCNTs.[53, 54]

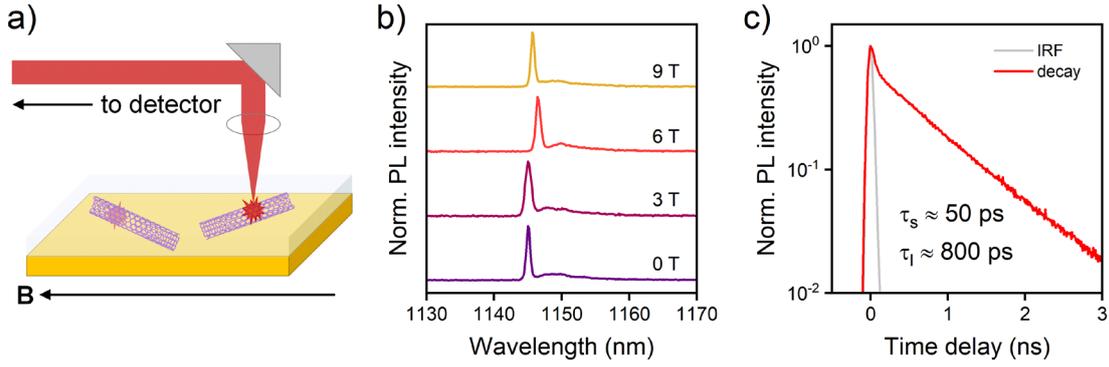

**Figure 6.** (a) Schematic of confocal magneto-PL measurement in Voigt geometry. (b) Magnetic field-dependent PL spectra of a single PTM-functionalized (6,5) nanotube (angle between tube axis and magnetic field: 60°). Note that minor shifts of the emission wavelength are due to spectral jitter and do not correlate with the applied field. (c) Low-temperature PL decay of the defect signal shown in (b) at zero field. All measurements were performed at 4 K.

Up to magnetic fields of 9 T, none of these defect signals displayed a significant shift or splitting within the resolution limit of the spectrometer (~300 µeV) and the characteristic spectral jitter (~2 meV) as exemplified in **Figure 6b** and more generally shown in **Supporting Information Figure S9**. This result is slightly surprising, because Kim *et al.* recently observed magnetic field-induced fine structure in the spectra of $sp^3$ defects.[48] Yet, peak splitting only occurred in a minority of nanotubes and the authors hypothesized that this was related to the requirement of nearly perfect energy level alignment between the bright $E_{11}*$ singlet and the band-edge triplet exciton. Within our limited statistics we cannot rule out that a fraction of nanotubes in our sample would show a similar effect, but based on the above estimation that $\Delta_{BT} \lesssim 38$ meV, it is likely that there is indeed no triplet state within just a few meV of the bright state in our PTM defects, which would explain the absence of a peak splitting. It is also worth noting that the triplet state, which is populated *via* EISC from an energetically close singlet state, usually does not show enhanced phosphorescence in magnetic fields, because the EISC rate also depends on the energy gap between the involved states[46] and the gap between the excited triplet and the singlet ground state is too large to result in significant triplet brightening.[42] Beyond that, the spin Zeeman effect that acts on the radical and splits its SOMO level is insufficient to induce a change of the defect PL, as the unpaired electron on the PTM is merely a spectator to the recombination of the defect-localized exciton with zero spin. Conceptually, the situation is analogous to that of defect-trapped trions in SWCNTs.[55] In both cases, the spin state does not change upon optical transition from the excited to the ground state, thus resulting in a vanishing spin Zeeman effect.

To gain further insights into the energetic position of the triplet state, we measured the PL decay of an individual PTM defect at 4 K (**Figure 6c**). Interestingly, we observed a biexponential decay characterized by an unusually[47, 56] fast component ($\tau_s \approx 50$ ps) with large amplitude ($A_s \approx 60\%$) and a slow component ($\tau_l \approx 800$ ps) typical for trapped exciton recombination at low temperature. Such a dominant fast process has never been reported for aryl defects with closed-shell substituents at low temperature. Note that multi-exciton effects can be excluded as the origin of the fast decay component, because the pump fluence (~$2 \times 10^{14}$ photons cm$^{-2}$ pulse$^{-1}$) was kept within the regime of linear PL response.[57] Since PET is too slow in our system while the observed $\tau_s$ matches the room temperature EISC rate well, we assign the fast decay component to EISC and conclude that the participating triplet state is energetically below the bright state. However, since RISC is prohibited at 4 K, the origin of $\tau_l$ is not obvious. We speculate that the surrounding insulating polymer matrix and cryogenic temperature could facilitate the trapping of charges on the PTM group, *e.g.*, after a PET event. This leads to electrochemical reduction of the radical to the closed-shell anion, in which EISC is switched off. In this case, $\tau_s$ corresponds to the decay time when the defect is in its native radical state, whereas $\tau_l$ represents the slower decay when the defect is trapped in its spin-paired anion state without EISC. Nevertheless, further work will be required to fully understand the PL dynamics of radical-functionalized defects in SWCNTs at cryogenic temperatures.

**CONCLUSION**

In summary, we have demonstrated the functionalization of monochiral semiconducting SWCNTs with stable organic radicals *via* simple diazonium chemistry. The presence of the radical leads to partial quenching of defect PL, which is likely due to a combination of a photo-induced electron transfer process and population transfer to triplet states enabled by radical-enhanced intersystem crossing. Consequently, radical-functionalized $sp^3$ defects are an interesting platform for future studies of the elusive triplet exciton manifold in carbon nanotubes employing spin-sensitive methods such as optically-detected magnetic resonance.[51, 58] Moreover, as pulsed EPR techniques can probe the distances between spin labels on the length scale of a few nanometers,[59] such defects could be a model system for investigating clustering, which is suspected to play an important part in both synthetic[60, 61] and optical[56] features of luminescent defects. Beyond that, the encapsulation of such radical-tailored SWCNTs in biocompatible surfactants[62] might even enable their use as metal-free contrast agents for magnetic resonance imaging[63, 64] complementary to *in vivo* near-infrared fluorescence imaging.[2] Finally, our functionalization approach should be easily transferrable to metallic SWCNTs, which show even higher reactivity towards diazonium reagents[65] and are commonly used in spin valves. In order to maximize the coupling between the radical and the itinerant electrons in such a system, the covalent linker should be as short as possible to place the radical close to the conducting channel.

**METHODS**

**Selective dispersion of (6,5) SWCNTs.** As described previously,[24] (6,5) SWCNTs were extracted from CoMoCAT raw material (Chasm Advanced Materials, SG65i-L63, 0.38 g L$^{-1}$) by selective wrapping with poly-[(9,9-dioctylfluorenyl-2,7-diyl)-*alt*-(6,6')-(2,2'-bipyridine)] (PFO-BPy, American Dye Source, M$_W$ = 40 kg mol$^{-1}$, 0.5 g L$^{-1}$) under shear force mixing (Silverson L2/Air, 10230 rpm, 72 h) in toluene. Aggregates and impurities were removed by two centrifugation steps (60000 *g* for 45 min each) and subsequent filtration (poly(tetrafluoroethylene) (PTFE) syringe filter, 5 µm pore size).

**SWCNT functionalization.** (6,5) SWCNTs were reacted with a series of diazonium compounds to introduce sp$^3$ aryl defects with either radical (PTM) or closed-shell substituents (PTMH, Br). Since the employed diazonium salts displayed different solubility characteristics, the functionalization procedure was adapted to this. The reactions between (6,5) SWCNTs and the tailored diazonium salts PTM-Dz and PTMH-Dz (see the **Supporting Information** for the synthesis[26] and characterization) were performed in tetrahydrofuran (THF), which was freshly distilled to remove water and any peroxide contaminations. First, the PFO-BPy-wrapped (6,5) SWCNTs were transferred to THF by passing the toluene dispersion through a PTFE membrane filter (Merck Millipore, JVWP, 0.1 µm pore size) to collect the SWCNTs, which were subsequently redispersed in a small volume of THF by bath sonication. To this dispersion, a solution of the diazonium salt (PTM-Dz or PTMH-Dz) in THF was added such that the (6,5) SWCNT concentration in the reaction mixture was 0.72 mg L$^{-1}$ (optical density of 0.4 cm$^{-1}$ at the E$_{11}$ transition) and the diazonium salt concentration 0.15 mmol L$^{-1}$. In contrast to that, the reaction with the commercial 4-bromobenzenediazonium tetrafluoroborate (Br-Dz, Sigma-Aldrich) was carried out according to our previously reported procedure[27] using 18-crown-6 as a phase-transfer agent in an 80:20 vol-% toluene/acetonitrile mixture, but without the addition of potassium acetate. The (6,5) SWCNT concentration was adjusted to 0.72 mg L$^{-1}$ and the Br-Dz concentration to 0.37 mmol L$^{-1}$ for moderate or 1.85 mmol L$^{-1}$ for high-density functionalization. All reactions proceeded in air at room temperature under dark conditions. After 16 hours, the functionalized SWCNTs were collected by vacuum filtration through a PTFE membrane filter (Merck Millipore, JVWP, 0.1 µm pore size) and washed with either THF (3 × 7 mL) or acetonitrile (10 mL) followed by toluene (5 mL), depending on the reaction medium. For analysis, the functionalized SWCNTs were redispersed in a small volume of toluene by bath sonication for 30 min.

**EPR spectroscopy.** EPR spectra were recorded at room temperature on a Bruker ESP 300 E spectrometer provided with a rectangular cavity T102 that works with an X band (9.5 GHz). The signal-to-noise ratio of spectra was increased by accumulation of scans using the F/Flock accessory to guarantee large field reproducibility. Precautions to avoid undesirable spectral distortion and line broadenings, such as those arising from microwave power saturation and magnetic field over-modulation, were also taken into account to improve sensitivity.

**Room temperature PL characterization.** PL spectra and time-correlated single photon counting (TCSPC) were measured for SWCNT dispersions using a home-built setup. The samples were excited by the spectrally filtered output of a picosecond-pulsed supercontinuum laser source (Fianium WhiteLase SC400) tuned to 700 nm to avoid any photochemical reactions of PTM-functionalized SWCNTs. Scattered laser light was blocked by appropriate longpass filters. The sample emission was dispersed by a grating spectrograph (Acton SpectraPro SP2358, 150 lines mm$^{-1}$) and detected by a liquid nitrogen cooled InGaAs line camera (Princeton Instruments OMA V:1024-1.7 LN). For TCSPC, the spectrally selected output of the spectrograph was focused onto a gated InGaAs/InP avalanche photodiode (Micro Photon Devices). Histograms of photon arrival times were created using a counting module (PicoHarp 300, PicoQuant) and reconvolution-fitted using the SymPhoTime 64 software. The instrument response function (IRF) was determined by the fast, instrument-limited decay of the $E_{11}$ emission from (6,5) SWCNTs.

**Low-temperature magneto-spectroscopy.** For PL microscopy on the single defect level, a low concentration of functionalized (6,5) SWCNTs (optical density of 0.005 cm$^{-1}$ at the $E_{11}$ transition) was blended into a solution of polystyrene (20 g L$^{-1}$) in toluene and spin-coated at 2000 rpm onto glass slides coated with 150 nm of gold. The sample was loaded into a home-built confocal microscope setup inside a closed-cycle magneto-cryostat (attoDRY1000, attocube systems) with a base temperature of 3.2 K and magnetic fields of up to 9 T. A customized holder was used to mount the sample in Voigt geometry with the magnetic field parallel and light propagation orthogonal to the sample surface. A wavelength-tunable Ti:sapphire laser (Coherent Mira) served as the excitation source and was tuned to 995 nm (continuous wave, 6 µW) to be in resonance with the $E_{11}$ transition of (6,5) SWCNTs. The PL from individual nanotubes was collected by an aspheric lens (Thorlabs 354330-B, N.A. = 0.68), dispersed by a grating spectrograph (Acton SpectraPro SP2500, 300 lines mm$^{-1}$) and recorded by a liquid nitrogen cooled InGaAs line camera (Princeton Instruments OMA V:1024-1.7 LN). For TCSPC, nanotubes were excited in pulsed mode (~2×10$^{14}$ photons cm$^{-2}$ pulse$^{-1}$) and the PL was directed to a superconducting single-photon detector (TCOPRS-CCR-SW-85, Scontel). The instrument response function (IRF) was measured on the attenuated laser signal.

**SUPPORTING INFORMATION**

Optical trap depth of PTM *vs.* PTMH defects. Highly functionalized closed-shell reference. Fine structure of the EPR signal. Broadening of the EPR signal. Proposed reaction scheme for open-shell to closed-shell conversion. Normalized spectra for irradiation experiment. Exclusion of energy transfer due to negligible spectral overlap. Model for PL decay dynamics of sp$^3$ defects with open-shell substituents. Scanning confocal PL map of single nanotubes. Magneto-PL spectra of single nanotubes. Synthesis and characterization of diazonium compounds including raw spectral data for PTMH-Dz.

**AUTHOR INFORMATION**

**Corresponding Author**

*E-mail: zaumseil@uni-heidelberg.de

**ORCID**

Felix J. Berger: 0000-0003-2834-0050

J. Alejandro de Sousa: 0000-0002-7948-8162

Jana Zaumseil: 0000-0002-2048-217X

Alexander Högele: 0000-0002-0178-9117

Núria Crivillers: 0000-0001-6538-2482

**ACKNOWLEDGMENT**

The authors thank Dr. Vega Lloveras for performing electron paramagnetic resonance measurements. This project has received funding from the European Research Council (ERC) under the European Union's Horizon 2020 research and innovation programme (Grant agreement No. 817494 "TRIFECTs"). S.Z. acknowledges funding from the Alexander von Humboldt Foundation, and A.H. from the European Research Council (ERC) under the Grant Agreement No. 772195 and the Deutsche Forschungsgemeinschaft (DFG, German Research Foundation) under Germany's Excellence Strategy EXC-2111-390814868. This work was also supported by the MICIU of Spain (PID2019-111682RB-I00), the Generalitat de Catalunya (2017-SGR-918) and the Severo Ochoa FUNFUTURE (CEX2019-000917-S). J. A. de S. is enrolled in the Materials Science PhD program of UAB. J. A. de S. thanks the FPI fellowship.

# REFERENCES

1. Luo, Y.; He, X.; Kim, Y.; Blackburn, J. L.; Doorn, S. K.; Htoon, H.; Strauf, S., Carbon Nanotube Color Centers in Plasmonic Nanocavities: A Path to Photon Indistinguishability at Telecom Bands. *Nano Lett.* **2019,** *19*, 9037-9044.

2. Mandal, A. K.; Wu, X.; Ferreira, J. S.; Kim, M.; Powell, L. R.; Kwon, H.; Groc, L.; Wang, Y.; Cognet, L., Fluorescent $sp^3$ Defect-Tailored Carbon Nanotubes Enable NIR-II Single Particle Imaging in Live Brain Slices at Ultra-Low Excitation Doses. *Sci. Rep.* **2020,** *10*, 5286.

3. Lee, Y.; Trocchia, S. M.; Warren, S. B.; Young, E. F.; Vernick, S.; Shepard, K. L., Electrically Controllable Single-Point Covalent Functionalization of Spin-Cast Carbon-Nanotube Field-Effect Transistor Arrays. *ACS Nano* **2018,** *12*, 9922-9930.

4. Durkop, T.; Getty, S. A.; Cobas, E.; Fuhrer, M. S., Extraordinary Mobility in Semiconducting Carbon Nanotubes. *Nano Lett.* **2004,** *4*, 35-39.

5. Hertel, T.; Himmelein, S.; Ackermann, T.; Stich, D.; Crochet, J., Diffusion Limited Photoluminescence Quantum Yields in 1-D Semiconductors: Single-Wall Carbon Nanotubes. *ACS Nano* **2010,** *4*, 7161-7168.

6. Brozena, A. H.; Kim, M.; Powell, L. R.; Wang, Y., Controlling the Optical Properties of Carbon Nanotubes with Organic Colour-Centre Quantum Defects. *Nat. Rev. Chem.* **2019,** *3*, 375-392.

7. Shiraki, T.; Miyauchi, Y.; Matsuda, K.; Nakashima, N., Carbon Nanotube Photoluminescence Modulation by Local Chemical and Supramolecular Chemical Functionalization. *Acc. Chem. Res.* **2020,** *53*, 1846-1859.

8. Gifford, B. J.; Kilina, S.; Htoon, H.; Doorn, S. K.; Tretiak, S., Controlling Defect-State Photophysics in Covalently Functionalized Single-Walled Carbon Nanotubes. *Acc. Chem. Res.* **2020,** *53*, 1791-1801.

9. Ghosh, S.; Bachilo, S. M.; Simonette, R. A.; Beckingham, K. M.; Weisman, R. B., Oxygen Doping Modifies Near-Infrared Band Gaps in Fluorescent Single-Walled Carbon Nanotubes. *Science* **2010,** *330*, 1656-1659.

10. Piao, Y.; Meany, B.; Powell, L. R.; Valley, N.; Kwon, H.; Schatz, G. C.; Wang, Y., Brightening of Carbon Nanotube Photoluminescence through the Incorporation of $sp^3$ Defects. *Nat. Chem.* **2013,** *5*, 840-845.

11. Kwon, H.; Furmanchuk, A.; Kim, M.; Meany, B.; Guo, Y.; Schatz, G. C.; Wang, Y., Molecularly Tunable Fluorescent Quantum Defects. *J. Am. Chem. Soc.* **2016,** *138*, 6878-6885.

12. He, X.; Hartmann, N. F.; Ma, X.; Kim, Y.; Ihly, R.; Blackburn, J. L.; Gao, W.; Kono, J.; Yomogida, Y.; Hirano, A.; Tanaka, T.; Kataura, H.; Htoon, H.; Doorn, S. K., Tunable Room-Temperature Single-Photon Emission at Telecom Wavelengths from $sp^3$ Defects in Carbon Nanotubes. *Nat. Photonics* **2017,** *11*, 577-582.

13. Gifford, B. J.; Kilina, S.; Htoon, H.; Doorn, S. K.; Tretiak, S., Exciton Localization and Optical Emission in Aryl-Functionalized Carbon Nanotubes. *J. Phys. Chem. C* **2018,** *122*, 1828-1838.

14. Kwon, H.; Kim, M.; Meany, B.; Piao, Y.; Powell, L. R.; Wang, Y., Optical Probing of Local pH and Temperature in Complex Fluids with Covalently Functionalized, Semiconducting Carbon Nanotubes. *J. Phys. Chem. C* **2015,** *119*, 3733-3739.

15. Shiraki, T.; Onitsuka, H.; Shiraishi, T.; Nakashima, N., Near Infrared Photoluminescence Modulation of Single-Walled Carbon Nanotubes Based on a Molecular Recognition Approach. *Chem. Commun.* **2016,** *52*, 12972-12975.

16. Onitsuka, H.; Fujigaya, T.; Nakashima, N.; Shiraki, T., Control of the Near Infrared Photoluminescence of Locally Functionalized Single-Walled Carbon Nanotubes *via* Doping by Azacrown-Ether Modification. *Chem. Eur. J.* **2018,** *24*, 9393-9398.

17. Mann, F. A.; Herrmann, N.; Opazo, F.; Kruss, S., Quantum Defects as a Toolbox for the Covalent Functionalization of Carbon Nanotubes with Peptides and Proteins. *Angew. Chem. Int. Ed. Engl.* **2020,** *59*, 17732-17738.

18. Schnee, M.; Besson, C.; Frielinghaus, R.; Lurz, C.; Kögerler, P.; Schneider, C. M.; Meyer, C., Quantum Transport in Carbon Nanotubes Covalently Functionalized with Magnetic Molecules. *Phys. Status Solidi B* **2016,** *253*, 2424-2427.

19. Ncube, S.; Coleman, C.; Strydom, A.; Flahaut, E.; de Sousa, A.; Bhattacharyya, S., Kondo Effect and Enhanced Magnetic Properties in Gadolinium Functionalized Carbon Nanotube Supramolecular Complex. *Sci. Rep.* **2018,** *8*, 8057.

20. Urdampilleta, M.; Klyatskaya, S.; Cleuziou, J. P.; Ruben, M.; Wernsdorfer, W., Supramolecular Spin Valves. *Nat. Mater.* **2011,** *10*, 502-506.

21. Ji, L.; Shi, J.; Wei, J.; Yu, T.; Huang, W., Air-Stable Organic Radicals: New-Generation Materials for Flexible Electronics? *Adv. Mater.* **2020,** *32*, 1908015.

22. Ratera, I.; Veciana, J., Playing with Organic Radicals as Building Blocks for Functional Molecular Materials. *Chem. Soc. Rev.* **2012,** *41*, 303-349.

23. Ballester, M., Inert Free Radicals (IFR): A Unique Trivalent Carbon Species. *Acc. Chem. Res.* **1985,** *18*, 380-387.

24. Graf, A.; Zakharko, Y.; Schießl, S. P.; Backes, C.; Pfohl, M.; Flavel, B. S.; Zaumseil, J., Large Scale, Selective Dispersion of Long Single-Walled Carbon Nanotubes with High Photoluminescence Quantum Yield by Shear Force Mixing. *Carbon* **2016,** *105*, 593-599.

25. Seber, G.; Muñoz, J.; Sandoval, S.; Rovira, C.; Tobias, G.; Mas-Torrent, M.; Crivillers, N., Synergistic Exploitation of the Superoxide Scavenger Properties of Reduced Graphene Oxide and a Trityl Organic Radical for the Impedimetric Sensing of Xanthine. *Adv. Mater. Interfaces* **2018,** *5*, 1701072.

26. Seber, G.; Rudnev, A. V.; Droghetti, A.; Rungger, I.; Veciana, J.; Mas-Torrent, M.; Rovira, C.; Crivillers, N., Covalent Modification of Highly Ordered Pyrolytic Graphite with a Stable Organic Free Radical by Using Diazonium Chemistry. *Chem. Eur. J.* **2017,** *23*, 1415-1421.

27. Berger, F. J.; Lüttgens, J.; Nowack, T.; Kutsch, T.; Lindenthal, S.; Kistner, L.; Müller, C. C.; Bongartz, L. M.; Lumsargis, V. A.; Zakharko, Y.; Zaumseil, J., Brightening of Long, Polymer-Wrapped Carbon Nanotubes by sp$^3$ Functionalization in Organic Solvents. *ACS Nano* **2019,** *13*, 9259-9269.

28. Kim, M.; Adamska, L.; Hartmann, N. F.; Kwon, H.; Liu, J.; Velizhanin, K. A.; Piao, Y.; Powell, L. R.; Meany, B.; Doorn, S. K.; Tretiak, S.; Wang, Y., Fluorescent Carbon Nanotube Defects Manifest Substantial Vibrational Reorganization. *J. Phys. Chem. C* **2016,** *120*, 11268-11276.

29. Zaka, M.; Ito, Y.; Wang, H.; Yan, W.; Robertson, A.; Wu, Y. A.; Rümmeli, M. H.; Staunton, D.; Hashimoto, T.; Morton, J. J. L.; Ardavan, A.; Briggs, G. A. D.; Warner, J. H., Electron Paramagnetic Resonance Investigation of Purified Catalyst-Free Single-Walled Carbon Nanotubes. *ACS Nano* **2010,** *4*, 7708-7716.

30. Weil, J. A.; Bolton, J. R., *Electron Paramagnetic Resonance*. John Wiley & Sons: Hoboken, 2007.

31. Reuel, N. F.; Dupont, A.; Thouvenin, O.; Lamb, D. C.; Strano, M. S., Three-Dimensional Tracking of Carbon Nanotubes within Living Cells. *ACS Nano* **2012,** *6*, 5420-5428.

32. Lohmann, S. H.; Trerayapiwat, K. J.; Niklas, J.; Poluektov, O. G.; Sharifzadeh, S.; Ma, X., sp$^3$-Functionalization of Single-Walled Carbon Nanotubes Creates Localized Spins. *ACS Nano* **2020,** *14*, 17675-17682.

33. He, X.; Velizhanin, K. A.; Bullard, G.; Bai, Y.; Olivier, J. H.; Hartmann, N. F.; Gifford, B. J.; Kilina, S.; Tretiak, S.; Htoon, H.; Therien, M. J.; Doorn, S. K., Solvent- and Wavelength-Dependent Photoluminescence Relaxation Dynamics of Carbon Nanotube sp$^3$ Defect States. *ACS Nano* **2018,** *12*, 8060-8070.

34. Hartmann, N. F.; Velizhanin, K. A.; Haroz, E. H.; Kim, M.; Ma, X.; Wang, Y.; Htoon, H.; Doorn, S. K., Photoluminescence Dynamics of Aryl sp$^3$ Defect States in Single-Walled Carbon Nanotubes. *ACS Nano* **2016,** *10*, 8355-8365.

35. Fox, M. A.; Gaillard, E.; Chen, C.-C., Photochemistry of Stable Free Radicals: The Photolysis of Perchlorotriphenylmethyl Radicals. *J. Am. Chem. Soc.* **1987,** *109*, 7088-7094.

36. Ballester, M.; Castañer, J.; Riera, J.; Pujadas, J.; Armet, O.; Onrubia, C.; Rio, J. A., Inert Carbon Free Radicals. 5. Perchloro-9-Phenylfluorenyl Radical Series. *J. Org. Chem.* **1984,** *49*, 770-778.

37. Weller, A., Photoinduced Electron Transfer in Solution: Exciplex and Radical Ion Pair Formation Free Enthalpies and Their Solvent Dependence. *Z. Phys. Chem.* **1982,** *133*, 93-98.

38. Miller, J. R.; Calcaterra, L. T.; Closs, G. L., Intramolecular Long-Distance Electron Transfer in Radical Anions. The Effects of Free Energy and Solvent on the Reaction Rates. *J. Am. Chem. Soc.* **1984,** *106*, 3047-3049.

39. Lewis, F. D.; Wu, T.; Zhang, Y.; Letsinger, R. L.; Greenfield, S. R.; Wasielewski, M. R., Distance-Dependent Electron Transfer in DNA Hairpins. *Science* **1997,** *277*, 673-676.

40. Shiraishi, T.; Shiraki, T.; Nakashima, N., Substituent Effects on the Redox States of Locally Functionalized Single-Walled Carbon Nanotubes Revealed by *in Situ* Photoluminescence Spectroelectrochemistry. *Nanoscale* **2017,** *9*, 16900-16907.

41. Souto, M.; Cui, H.; Pena-Alvarez, M.; Baonza, V. G.; Jeschke, H. O.; Tomic, M.; Valenti, R.; Blasi, D.; Ratera, I.; Rovira, C.; Veciana, J., Pressure-Induced Conductivity in a Neutral Nonplanar Spin-Localized Radical. *J. Am. Chem. Soc.* **2016,** *138*, 11517-11525.

42. Wang, Z.; Zhao, J.; Barbon, A.; Toffoletti, A.; Liu, Y.; An, Y.; Xu, L.; Karatay, A.; Yaglioglu, H. G.; Yildiz, E. A.; Hayvali, M., Radical-Enhanced Intersystem Crossing in New Bodipy Derivatives and Application for Efficient Triplet-Triplet Annihilation Upconversion. *J. Am. Chem. Soc.* **2017,** *139*, 7831-7842.

43. Colvin, M. T.; Giacobbe, E. M.; Cohen, B.; Miura, T.; Scott, A. M.; Wasielewski, M. R., Competitive Electron Transfer and Enhanced Intersystem Crossing in Photoexcited Covalent Tempo-Perylene-3,4:9,10-Bis(dicarboximide) Dyads: Unusual Spin Polarization Resulting from the Radical-Triplet Interaction. *J. Phys. Chem. A* **2010,** *114*, 1741-1748.

44. Dyar, S. M.; Margulies, E. A.; Horwitz, N. E.; Brown, K. E.; Krzyaniak, M. D.; Wasielewski, M. R., Photogenerated Quartet State Formation in a Compact Ring-Fused Perylene-Nitroxide. *J. Phys. Chem. B* **2015,** *119*, 13560-13569.

45. Giacobbe, E. M.; Mi, Q.; Colvin, M. T.; Cohen, B.; Ramanan, C.; Scott, A. M.; Yeganeh, S.; Marks, T. J.; Ratner, M. A.; Wasielewski, M. R., Ultrafast Intersystem Crossing and Spin Dynamics of Photoexcited Perylene-3,4:9,10-Bis(dicarboximide) Covalently Linked to a Nitroxide Radical at Fixed Distances. *J. Am. Chem. Soc.* **2009,** *131*, 3700-3712.

46. Yeganeh, S.; Wasielewski, M. R.; Ratner, M. A., Enhanced Intersystem Crossing in Three-Spin Systems: A Perturbation Theory Treatment. *J. Am. Chem. Soc.* **2009,** *131*, 2268-2273.

47. Kim, Y.; Velizhanin, K. A.; He, X.; Sarpkaya, I.; Yomogida, Y.; Tanaka, T.; Kataura, H.; Doorn, S. K.; Htoon, H., Photoluminescence Intensity Fluctuations and Temperature-Dependent Decay Dynamics of Individual Carbon Nanotube $sp^3$ Defects. *J. Phys. Chem. Lett.* **2019,** *10*, 1423-1430.

48. Kim, Y.; Goupalov, S. V.; Weight, B. M.; Gifford, B. J.; He, X.; Saha, A.; Kim, M.; Ao, G.; Wang, Y.; Zheng, M.; Tretiak, S.; Doorn, S. K.; Htoon, H., Hidden Fine Structure of Quantum Defects Revealed by Single Carbon Nanotube Magneto-Photoluminescence. *ACS Nano* **2020,** *14*, 3451-3460.

49. Ishii, A.; Machiya, H.; Kato, Y. K., High Efficiency Dark-to-Bright Exciton Conversion in Carbon Nanotubes. *Phys. Rev. X* **2019,** *9*, 041048.

50. Gokus, T.; Cognet, L.; Duque, J. G.; Pasquali, M.; Hartschuh, A.; Lounis, B., Mono- and Biexponential Luminescence Decays of Individual Single-Walled Carbon Nanotubes. *J. Phys. Chem. C* **2010,** *114*, 14025-14028.

51. Stich, D.; Späth, F.; Kraus, H.; Sperlich, A.; Dyakonov, V.; Hertel, T., Triplet–Triplet Exciton Dynamics in Single-Walled Carbon Nanotubes. *Nat. Photonics* **2013,** *8*, 139-144.

52. Lin, C. W.; Bachilo, S. M.; Weisman, R. B., Delayed Fluorescence from Carbon Nanotubes through Singlet Oxygen-Sensitized Triplet Excitons. *J. Am. Chem. Soc.* **2020,** *142*, 21189–21196.

53. Sarpkaya, I.; Ahmadi, E. D.; Shepard, G. D.; Mistry, K. S.; Blackburn, J. L.; Strauf, S., Strong Acoustic Phonon Localization in Copolymer-Wrapped Carbon Nanotubes. *ACS Nano* **2015,** *9*, 6383-6393.

54. He, X.; Gifford, B. J.; Hartmann, N. F.; Ihly, R.; Ma, X.; Kilina, S. V.; Luo, Y.; Shayan, K.; Strauf, S.; Blackburn, J. L.; Tretiak, S.; Doorn, S. K.; Htoon, H., Low-Temperature Single Carbon Nanotube Spectroscopy of sp$^3$ Quantum Defects. *ACS Nano* **2017,** *11*, 10785-10796.

55. Kwon, H.; Kim, M.; Nutz, M.; Hartmann, N. F.; Perrin, V.; Meany, B.; Hofmann, M. S.; Clark, C. W.; Htoon, H.; Doorn, S. K.; Högele, A.; Wang, Y., Probing Trions at Chemically Tailored Trapping Defects. *ACS Cent. Sci.* **2019,** *5*, 1786-1794.

56. Nutz, M.; Zhang, J.; Kim, M.; Kwon, H.; Wu, X.; Wang, Y.; Högele, A., Photon Correlation Spectroscopy of Luminescent Quantum Defects in Carbon Nanotubes. *Nano Lett.* **2019,** *19*, 7078-7084.

57. Harrah, D. M.; Schneck, J. R.; Green, A. A.; Hersam, M. C.; Ziegler, L. D.; Swan, A. K., Intensity-Dependent Exciton Dynamics of (6,5) Single-Walled Carbon Nanotubes: Momentum Selection Rules, Diffusion, and Nonlinear Interactions. *ACS Nano* **2011,** *5*, 9898-9906.

58. Palotas, J.; Negyedi, M.; Kollarics, S.; Bojtor, A.; Rohringer, P.; Pichler, T.; Simon, F., Incidence of Quantum Confinement on Dark Triplet Excitons in Carbon Nanotubes. *ACS Nano* **2020,** *14*, 11254-11261.

59. Schiemann, O.; Cekan, P.; Margraf, D.; Prisner, T. F.; Sigurdsson, S. T., Relative Orientation of Rigid Nitroxides by Peldor: Beyond Distance Measurements in Nucleic Acids. *Angew. Chem. Int. Ed. Engl.* **2009,** *48*, 3292-3295.

60. Zhang, Y.; Valley, N.; Brozena, A. H.; Piao, Y.; Song, X.; Schatz, G. C.; Wang, Y., Propagative Sidewall Alkylcarboxylation That Induces Red-Shifted Near-IR Photoluminescence in Single-Walled Carbon Nanotubes. *J. Phys. Chem. Lett.* **2013,** *4*, 826-830.

61. Danné, N.; Kim, M.; Godin, A. G.; Kwon, H.; Gao, Z.; Wu, X.; Hartmann, N. F.; Doorn, S. K.; Lounis, B.; Wang, Y.; Cognet, L., Ultrashort Carbon Nanotubes That Fluoresce Brightly in the Near-Infrared. *ACS Nano* **2018,** *12*, 6059-6065.

62. Gao, Z.; Varela, J. A.; Groc, L.; Lounis, B.; Cognet, L., Toward the Suppression of Cellular Toxicity from Single-Walled Carbon Nanotubes. *Biomater. Sci.* **2016,** *4*, 230-244.

63. Marangon, I.; Ménard-Moyon, C.; Kolosnjaj-Tabi, J.; Béoutis, M. L.; Lartigue, L.; Alloyeau, D.; Pach, E.; Ballesteros, B.; Autret, G.; Ninjbadgar, T.; Brougham, D. F.; Bianco, A.; Gazeau, F., Covalent Functionalization of Multi-Walled Carbon Nanotubes with a Gadolinium Chelate for Efficient T$_1$-Weighted Magnetic Resonance Imaging. *Adv. Funct. Mater.* **2014,** *24*, 7173-7186.

64.     Rajca, A.; Wang, Y.; Boska, M.; Paletta, J. T.; Olankitwanit, A.; Swanson, M. A.; Mitchell, D. G.; Eaton, S. S.; Eaton, G. R.; Rajca, S., Organic Radical Contrast Agents for Magnetic Resonance Imaging. *J. Am. Chem. Soc.* **2012,** *134*, 15724-15727.

65.     Strano, M. S.; Dyke, C. A.; Usrey, M. L.; Barone, P. W.; Allen, M. J.; Shan, H.; Kittrell, C.; Hauge, R. H.; Tour, J. M.; Smalley, R. E., Electronic Structure Control of Single-Walled Carbon Nanotube Functionalization. *Science* **2003,** *301*, 1519-1522.

# Supporting Information

# Interaction of Luminescent Defects in Carbon Nanotubes with Covalently Attached Stable Organic Radicals

*Felix J. Berger[1,2], J. Alejandro de Sousa[3,4], Shen Zhao[5,6], Nicolas F. Zorn[1,2], Abdurrahman Ali El Yumin[1,2], Aleix Quintana García[3], Simon Settele[1], Alexander Högele[5,6], Núria Crivillers[3] and Jana Zaumseil[1,2]\**

[1]Institute for Physical Chemistry, Universität Heidelberg, 69120 Heidelberg, Germany

[2]Centre for Advanced Materials, Universität Heidelberg, 69120 Heidelberg, Germany

[3]Institut de Ciència de Materials de Barcelona (ICMAB-CSIC), Campus UAB, 08193 Bellaterra, Spain

[4]Laboratorio de Electroquímica, Departamento de Química, Facultad de Ciencias, Universidad de los Andes, 5101 Mérida, Venezuela

[5]Faculty of Physics, Munich Quantum Center and Center for NanoScience (CeNS), Ludwig-Maximilians-Universität München, 80539 München, Germany

[6]Munich Center for Quantum Science and Technology (MCQST), 80799 München, Germany

Corresponding Author

\* E-mail: zaumseil@uni-heidelberg.de

# TABLE OF CONTENTS

## Optical trap depth of PTM *vs.* PTMH defects

In order to extract the optical trap depths of PTM and PTMH defects with reasonable accuracy despite the low defect density, all PL spectra were normalized to the $E_{11}$ feature (see **Figure 2a**, main text) and the pristine spectrum was subtracted from the spectra of functionalized SWCNTs. Thereby, the residual $E_{11}$-associated phonon sideband signal is removed and the pure defect emission spectra were obtained (**Figure S1**). The optical trap depth is defined as the emission energy difference between the $E_{11}$ peak and the $E_{11}^*$ peak. The optical trap depths of PTM (172 meV) and PTMH defects (168 meV) are very similar.

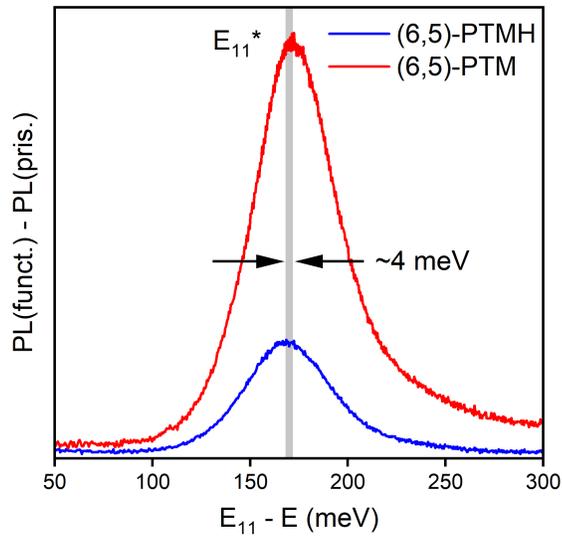

**Figure S1.** Difference PL spectra of PTM- and PTMH-functionalized (6,5) SWCNTs plotted on an energy scale relative to the $E_{11}$ peak emission energy.

## Highly functionalized closed-shell reference

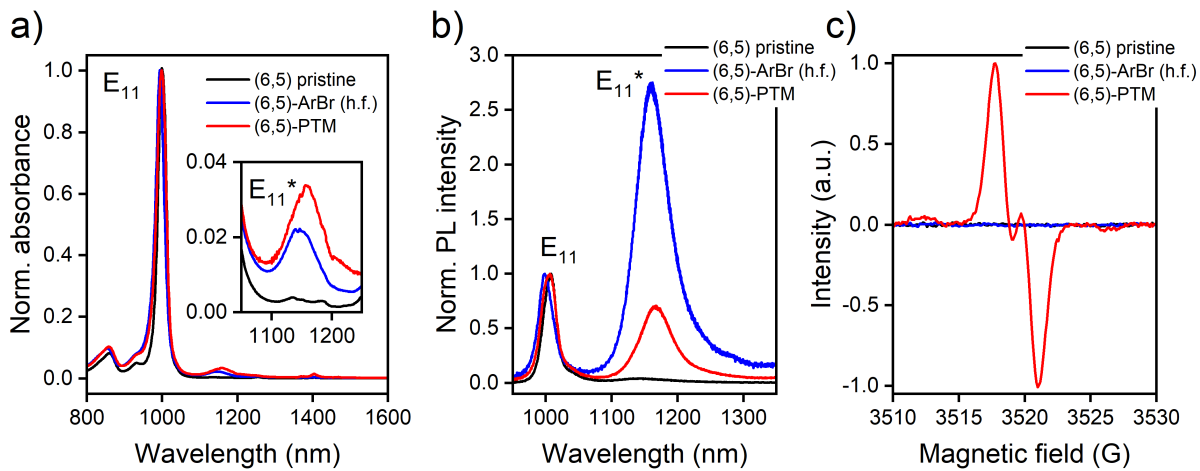

**Figure S2.** Characterization of pristine (6,5) SWCNTs, PTM-functionalized (6,5) SWCNTs and ArBr-functionalized (6,5) SWCNTs with high defect density (h.f.) in toluene dispersions. (a) Normalized absorption spectra. The absorption band at ~1160 nm indicates $sp^3$ defect formation.[1] The optical density at the $E_{11}$* band is comparable for PTM- and ArBr-functionalized SWCNTs and thus, their defect densities are similar. (b) Normalized PL spectra. Note that the large difference in the $E_{11}$*/$E_{11}$ PL intensity ratio originates from PTM radical-induced PL quenching as demonstrated in **Figure 3** and **Figure 4** of the main text. (c) EPR spectra. Neither pristine SWCNTs, nor SWCNTs with closed-shell functional groups (ArBr) give rise to an EPR signal at room temperature. The PTM-functionalized SWCNTs, on the other hand, clearly show the signature of the PTM-styryl unit. Hence, the EPR signal in PTM-tailored SWCNTs arises from an unpaired electron localized on the PTM moiety and not on the nanotube itself.

## Fine structure of the EPR signal

As the radical spin density is delocalized over several atoms of the PTM moiety, the magnetic moments of the involved nuclei can potentially couple to the radical spin and lead to a peak splitting in the EPR spectrum. For the PTM-styryl unit that is attached to the SWCNT *via* the diazonium reagent PTM-Dz (structure shown in **Figure S3**), the $^{13}$C and $^{1}$H nuclei give rise to significant coupling. The radical-bearing carbon atom and the aromatic carbon atoms show the largest couplings (J ~ 30 G and J ~ 10-15 G, respectively), but since $^{13}$C has a low natural abundance (1.1 %) the corresponding EPR signals are weak. In addition, one of the $^{1}$H atoms on the vinylene unit is of the correct symmetry to couple to the radical (indicated in **Figure S3**). The coupling is small (J ~ 2 G) but the EPR signals are quite prominent due to the large abundance of $^{1}$H (99.9 %).

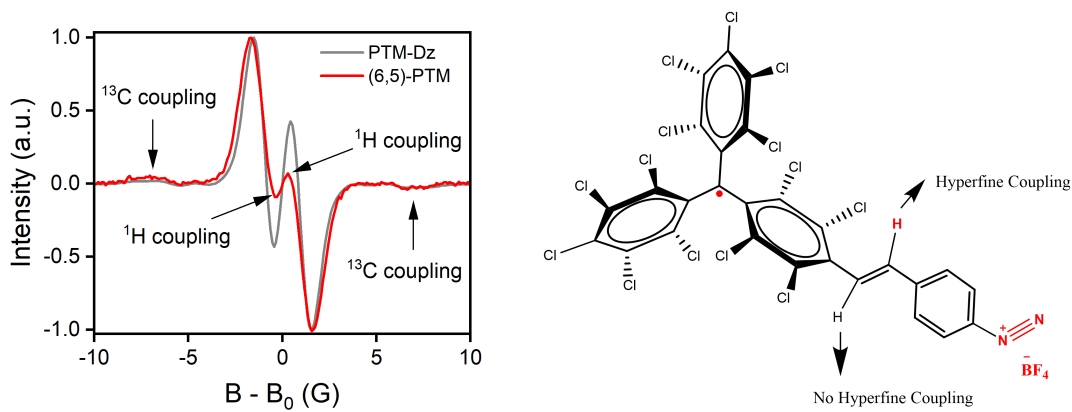

**Figure S3.** EPR spectra of the PTM diazonium salt (gray) and the PTM-functionalized (6,5) SWCNTs (red). For clarity, the spectra are plotted as a function of $B - B_0$, where $B_0$ is the field at the center of the signal. The $^{1}$H nucleus with appropriate symmetry to couple to the radical is indicated in the molecular structure of the PTM diazonium salt.

**Broadening of the EPR signal**

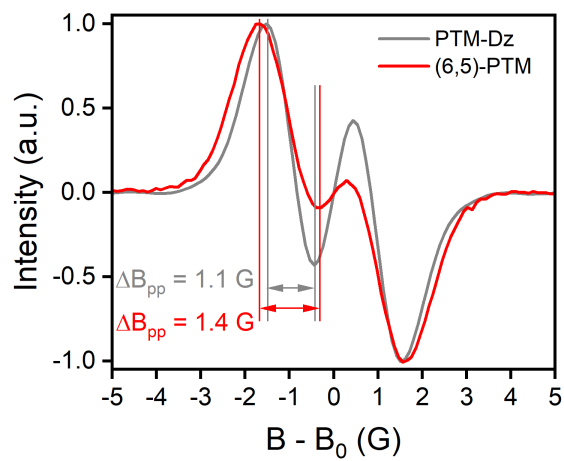

**Figure S4.** A zoom on the EPR spectra of the PTM diazonium salt (gray) and the PTM-functionalized (6,5) SWCNTs (red). For clarity, the spectra are plotted as a function of $B - B_0$, where $B_0$ is the field at the center of the signal. The linewidths ($\Delta B_{pp}$) are indicated in the graph. As a result of the increased linewidth, the apparent intensity of the $^1H$ coupling signals decreases due to cancellation of positive and negative contributions in the first derivative spectrum.

# Proposed reaction scheme for open-shell to closed-shell conversion

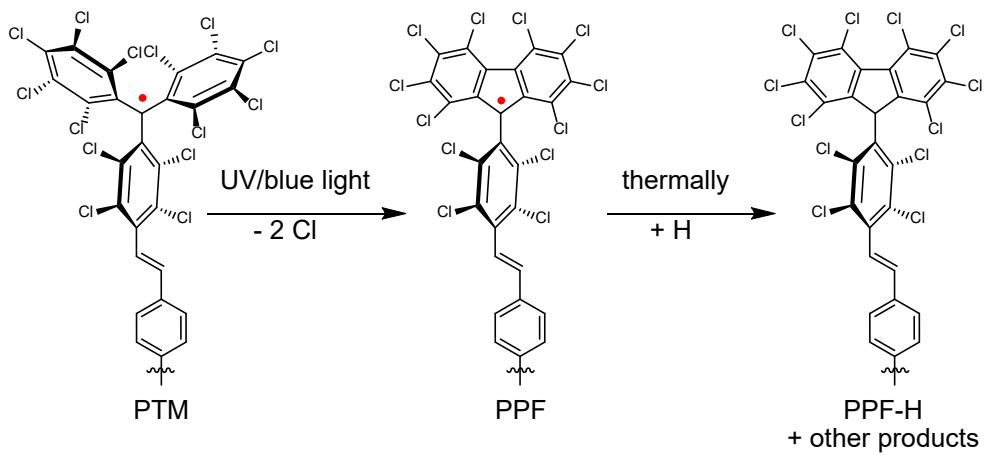

**Figure S5.** Proposed reaction pathway for the UV light-induced conversion of open-shell PTM defects to closed-shell species. The formation of the perchlorophenylfluorenyl (PPF) radical under UV irradiation is well documented.[2, 3] Due to the planar structure of the fluorene unit, the PPF radical is substantially less stable than the PTM radical and it should readily react with other molecules in solution. As shown in the scheme, hydrogen abstraction from the solvent is a plausible decomposition pathway for the PPF radical, but other reactions are possible too. Note that the extremely low concentration (~nmol/L) of $sp^3$ defects in the sample dispersion did not allow for the determination of the precise molecular structure of the photolysis products. However, the absence of an EPR signal from the irradiated sample (see **Figure 4d**, main text) confirms the closed-shell character of the substituents after irradiation.

**Normalized spectra for irradiation experiment**

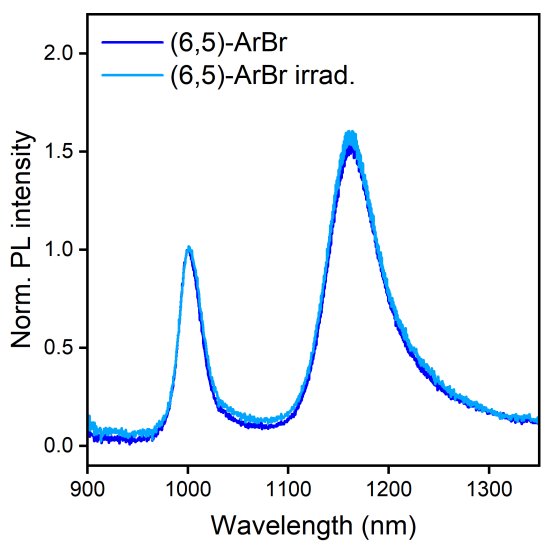

**Figure S6.** Normalized PL spectra of ArBr-functionalized (6,5) SWCNTs before and after UV irradiation. The identical spectral shapes show that common $sp^3$ defects are not affected by UV irradiation and further corroborates that the changes observed for PTM defects are due to the photochemical transformation of the PTM radical (see **Figure 4** in the main text).

# Exclusion of energy transfer due to negligible spectral overlap

The rates of Förster- and Dexter-type energy transfer[4, 5] from a donor (the $sp^3$ defect) to an acceptor (the PTM radical) depend on the overlap between the emission spectrum of the donor and the absorption spectrum of the acceptor. Since the density of covalently attached PTM radicals is low (otherwise the $E_{11}$ and $E_{11}^*$ PL from the SWCNT would both be quenched), the absorption due to nanotube-bound PTM groups is below our detection limit. For this reason, we employ two molecular PTM derivatives (triazene PTM-Taz and diazonium salt PTM-Dz, see the **Synthesis** section for molecular structures) as references. The strong absorption at ~385 nm is characteristic of PTM radicals. As shown in **Figure S7**, the spectral overlap between the $E_{11}^*$ emission and PTM absorption is negligible and therefore, energy transfer from the $sp^3$ defect to the PTM moiety can be excluded.

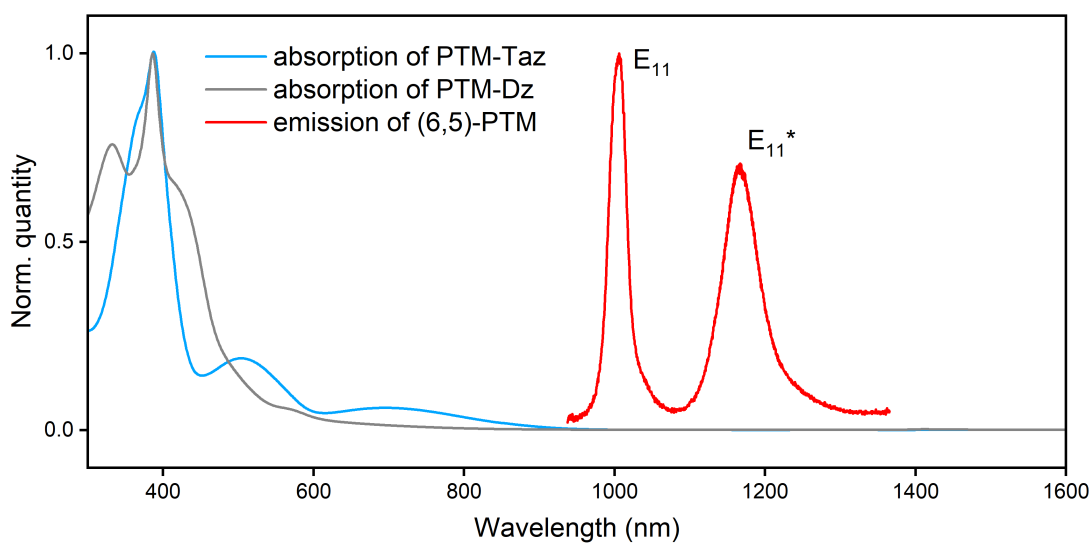

**Figure S7.** Absorption spectra of the molecular PTM radical derivatives PTM-Taz and PTM-Dz (refer to the **Synthesis** section below for their molecular structures) measured in tetrahydrofuran and emission spectrum of PTM-functionalized (6,5) SWCNTs measured in toluene.

# Model for PL decay dynamics of sp³ defects with open-shell substituents

This section complements the discussion of PET and EISC as PL quenching mechanisms in PTM-functionalized defects and explains the reasoning behind the model presented in **Figure 5b** of the main text. The key aspects will be addressed individually below.

**Energy level ordering of the B, D and T state.** Several studies found evidence that in sp³ defects the parity-forbidden dark singlet exciton (D) is a few meV above the bright singlet exciton (B).[6-8] This is in contrast to the band-edge exciton manifold, where the lowest singlet state is dark.[9] Concerning the energy of the triplet (T) state involved in the EISC process, the main argument for placing it below the B state comes from our low-temperature PL lifetime measurement (**Figure 6c**). We observe a fast decay component (50 ps) with a large amplitude (60 %), which is highly unusual[7,10] for sp³ defects at cryogenic temperature. PET is obviously too slow to account for this rate, because we find $k_{PET}$ = (500 ps)$^{-1}$ at room temperature and, being a thermally activated process, PET should slow down even further upon cooling. Therefore, we attribute the fast component to EISC from the bright state to a triplet state. Since the thermal energy of ~0.3 meV at 4 K precludes thermally activated EISC to higher energy states, we conclude that the participating T state is below the B state.

**Requirement of small $\Delta_{BT}$ and simultaneous PET and EISC.** Any process that leads to irreversible loss of bright (B) state population will enhance the rates $\tau_l^{-1}$ and $\tau_s^{-1}$ equally as it can occur at any time after exciton trapping.[11] Such a process could be PET or irreversible EISC to a low-lying triplet (T) state, because in that case, an energy gap $\Delta_{BT}$ of much more than 25 meV would prevent thermally activated reverse intersystem crossing (RISC). If one of these processes with hypothetical rate $k_Q$ was solely responsible for the observed quenching, the rates $\tau_l^{-1}$ and $\tau_s^{-1}$ should be enhanced equally according to:

$$\tau_l^{-1}(\text{PTM defect}) = \tau_l^{-1}(\text{closed-shell reference}) + k_Q$$

$$\tau_s^{-1}(\text{PTM defect}) = \tau_s^{-1}(\text{closed-shell reference}) + k_Q$$

However, we find experimentally that the discrepancies between PTM defects and "irradiated PTM" (closed-shell) defects are markedly different in $\tau_l^{-1}$ and $\tau_s^{-1}$. Using data from **Table 1** of the main text:

$$\tau_l^{-1}(\text{PTM defect}) - \tau_l^{-1}(\text{PTM irrad.}) = 2.1 \times 10^9 s^{-1}$$

$$\tau_s^{-1}(\text{PTM defect}) - \tau_s^{-1}(\text{PTM irrad.}) = 1.9 \times 10^{10} s^{-1}$$

Thus, there is a mismatch of almost one order of magnitude in the rates $k_Q$ extracted from $\tau_l^{-1}$ and $\tau_s^{-1}$. Therefore, a single quenching mechanism (PET or irreversible EISC) with rate $k_Q$ cannot account for the observed lifetime changes as long as the lifetime components are assigned to redistribution of exciton population ($\tau_s$) and exciton recombination ($\tau_l$) as is done in the established model introduced by Hartmann et al.[12] and He et al.[11]

The only way to account for the unequal changes in $\tau_l^{-1}$ and $\tau_s^{-1}$ is to invoke reversible EISC by assuming that $\Delta_{BT}$ is not much larger than 25 meV, such that RISC is feasible at room temperature. (The literature on thermally activated delayed fluorescence (TADF) emitters usually assumes that RISC is feasible at room temperature as long as the singlet-triplet splitting is less than 100 meV.[13]) In that case, the reversible population exchange between the B and the T state is responsible for the strong change in $\tau_s^{-1}$, while it does not affect the exciton recombination ($\tau_l^{-1}$). Hence, the missing term, i.e., the enhanced recombination rate, is assigned to a PET process according to $\tau_l^{-1}$ = k + $k_{PET}$. In summary, a combination of both mechanisms is required to adequately describe the PL data (see **Table 1** in the main text).

**Dominant role of the T state in the population redistribution process.** In radical-functionalized defects, both the D and the T state should be able to exchange population with the B state. Nevertheless, $\tau_s^{-1}$ is strongly enhanced in such defects, which suggests that the B-T equilibration ($k_{BT}$) is substantially faster than the B-D equilibration ($k_{BD}$) and dominates the observed $\tau_s^{-1}$. This appears reasonable given that the T state has the lowest energy of the involved excited states and thus, the B state population (which is probed in the time-resolved PL measurement) should be preferentially redistributed to the T state instead of the D state.

**$A_l$ as a measure of bright state population in equilibrium.** He et al.[11] pointed out that in the slow recombination limit ($\tau_s^{-1} \gg \tau_l^{-1}$) the normalized amplitude of the long decay component ($A_l$) may be taken as a measure of bright state population in equilibrium ($p_B^{eq}$) with respect to its initial population immediately after trapping ($p_B^0$), i.e. $A_l = \frac{p_B^{eq}}{p_B^0} = \frac{p_B^{eq}}{1-p_D^0-p_T^0}$.

**Relative contributions of PET and EISC to the PL quenching.** The contribution of PET to the quenching is found by evaluating the change in the amplitude-averaged PL lifetime ($\tau_{amp\text{-}av}$) resulting from the enhanced recombination rate: $\tau_{l,PET}^{-1} = k + k_{PET} = \tau_l^{-1} + k_{PET}$. In that case, the quenching factor ($QF_{PET}$) is defined as

$$QF_{PET} = \frac{\tau_{amp-av,PET}^{-1}}{\tau_{amp-av}^{-1}} = \frac{A_l(\tau_l^{-1} + k_{PET}) + A_s(\tau_l^{-1} + k_{PET} + k_{BD})}{A_l\tau_l^{-1} + A_s(\tau_l^{-1} + k_{BD})} = 1 + \frac{k_{PET}}{\tau_{amp-av}^{-1}}$$

where we used the amplitude normalization, *i.e.* $A_l + A_s = 1$. The resulting expression is the well-known Stern-Volmer equation[4] with $k_{PET}$ as the rate of quenching. Using the amplitude-averaged PL lifetime determined for the closed-shell irradiated PTM defects, $\tau_{amp-av}$ = (121 ps)$^{-1}$, and the extracted $k_{PET}$ = (500 ps)$^{-1}$, we obtain a quenching factor $QF_{PET}$ = 1.24. Hence, quenching purely due to PET can only account for an $E_{11}$* PL intensity reduction to $\frac{1}{1.24} \approx 0.81$ of the closed-shell reference value. As the $E_{11}$* PL is actually quenched to the level of $\approx 0.42$ in PTM defects, we estimate that about 1/3 of the intensity drop may be due to PET, whereas 2/3 are a consequence of EISC.

## Scanning confocal PL map of single nanotubes

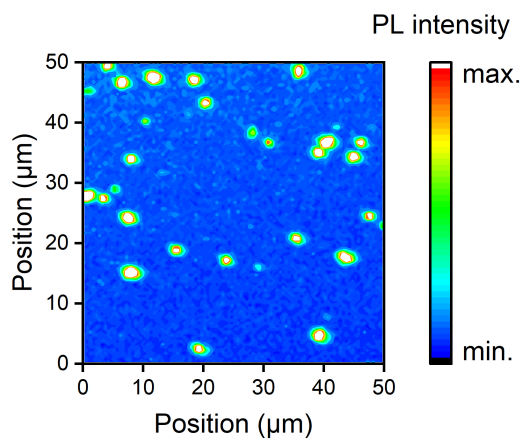

**Figure S8.** Raster-scanning confocal PL map of individual PTM-functionalized (6,5) SWCNTs embedded in a polystyrene layer on a gold-coated glass substrate. Brightness variations between single nanotubes are intensified by the match or mismatch between the random nanotube orientation and the linear polarization of the excitation laser (cw, 785 nm, 10 µW).

## Magneto-PL spectra of single nanotubes

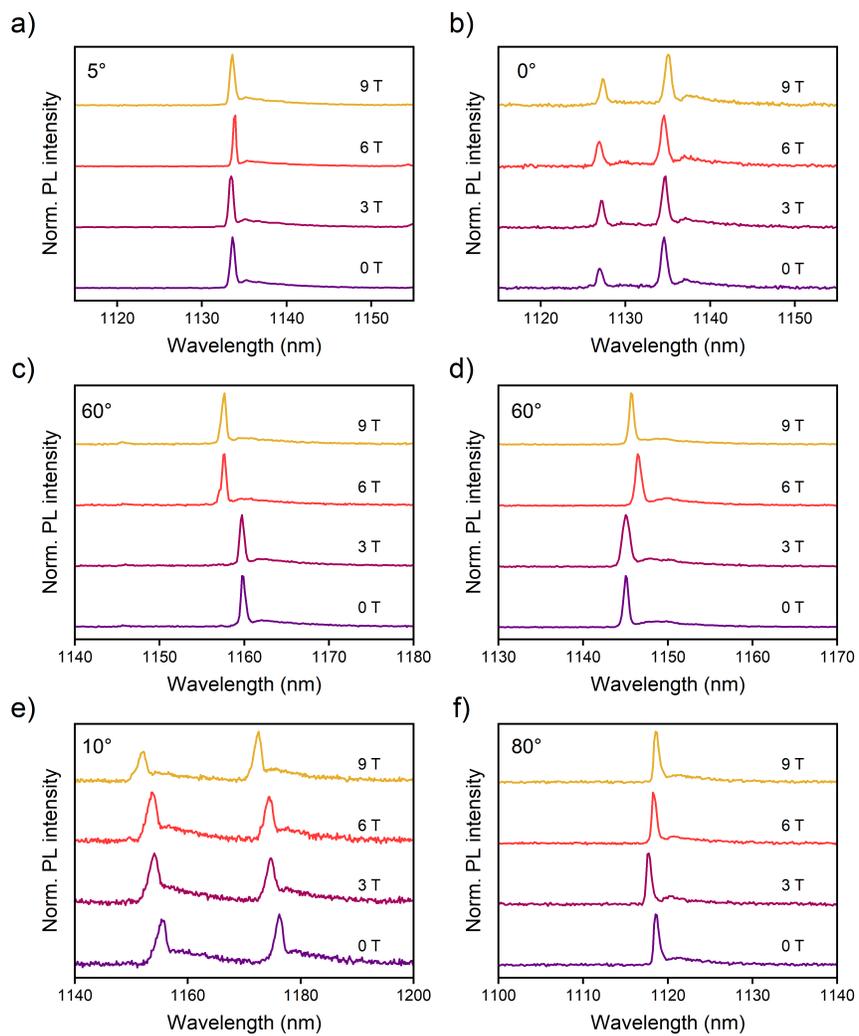

**Figure S9.** Magnetic field-dependent PL spectra of individual PTM-functionalized (6,5) SWCNTs (a-f). The angle between the nanotube axis and the in-plane magnetic field is provided in the graph. All measurements were performed at 4 K.

# Synthesis and characterization of diazonium compounds

## Chemicals used in synthesis

Aqueous tetrabutylammonium hydroxide 54-56 %, tetrabutylammonium hexafluorophosphate 98 %, potassium tert-butoxide anhydrous 99 %, sodium nitrite 97 % and 4-aminobenzaldehyde (Sigma-Aldrich). *p*-chloranil 99 %, sodium carbonate 99 % and hydrochloric acid 37 % (Panreac), tetrafluoroboric acid solution 50 % (Abcr). All solvents were of HPLC grade. Toluene and THF were dried using sodium and benzophenone as indicator before use; acetonitrile and dichloromethane were dried using calcium hydride before use. The purification of the synthesized compounds was carried out using Carlo Erba silica gel (60, particle size 35-70).

## Equipment used for characterization

**UV/vis absorption spectroscopy.** UV/vis absorption spectra were recorded with a JASCO V-780 UV/visible/NIR spectrophotometer. Quartz cuvettes with an optical path of 1 cm were used in all measurements.

**Nuclear magnetic resonance (NMR) spectroscopy.** The $^1$H-NMR and $^{13}$C-NMR spectra were acquired with a Bruker Avance-II+ (600MHz) spectrometer. The calibration was made using residual undeuterated acetone ($\delta(^1H)$ = 2.05 ppm; $\delta(^{13}C)$ = 29.84 ppm) as the internal reference. The data analysis was carried out with MestReNova software (MestReLab Research S. L.). The following abbreviations were used to designate multiplicities: s = singlet, d = doublet, m = multiplet.

**Infrared (IR) spectroscopy.** IR spectra were recorded with a Jasco 4700 Fourier transform (FTIR) spectrometer with a diamond ATR accessory in the spectral range from 400-4000 cm$^{-1}$ (resolution 4 cm$^{-1}$, 64 scans).

**Electrochemical characterization.** Measurements were performed with an AUTOLAB 204 potentiostat equipped with NOVA 2.3 software. Glassy carbon (exposed area of 0.28 cm$^2$) was used as the working electrode. A Pt mesh was used as the counter electrode, Ag wire was used as a pseudo-reference electrode and the ferrocene redox pair as an internal reference. The cell was kept under argon atmosphere and a 0.2 mol L$^{-1}$ solution of tetrabutylammonium hexafluorophosphate (TBAPF$_6$) in dry dichloromethane (DCM) was used as the electrolyte.

**Synthesis**

The PTM radical diazonium salt (**PTM-Dz**) and PTMH diazonium salt (**PTMH-Dz**) were synthesized *via* the corresponding triazenes (here referred to as **PTM-Taz** and **PTMH-Taz**). The compound **PTMH-Taz** was synthesized as described previously.[14] In this work, an improved procedure for radical generation from the **PTMH-Taz** was developed, which makes use of *p*-chloranil as the oxidizing agent instead of silver nitrate. This procedure is reported in the following.

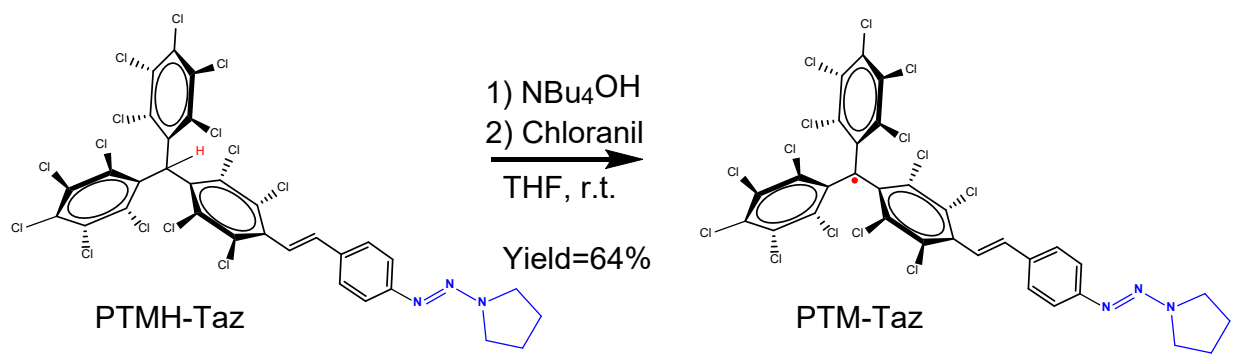

**Synthesis of PTM-Taz from PTMH-Taz.** The synthesis of **PTM-Taz** was carried out in a laboratory equipped with red light to avoid the decomposition of the radical species in solution. Tetrabutylammonium hydroxide (55 % aqueous, 74 µL, 0.16 mmol) was added to a solution of compound **PTMH-Taz** (101 mg, 0.11 mmol) in dry THF (6 mL) and the solution was stirred at room temperature. The formation of the PTM anion was monitored by UV/vis spectroscopy. When the deprotonation was complete (after about 30 min), *p*-chloranil was added (41.3 mg, 0.17 mmol) and the oxidation of the PTM anion to the radical was followed by UV/vis spectroscopy. When the oxidation was complete (after about 120 min) the solvent was evaporated under vacuum and the crude was purified by column chromatography (silica gel (normal phase), petroleum ether/ethyl acetate 90:10, column dimensions: L = 20 cm, ϕ = 3 cm). Compound **PTM-Taz** was obtained as a dark violet powder (64.8 mg, 64 % yield). ATR-IR: ν (cm$^{-1}$) = 2960 (C-H), 2921 (C-H), 2867 (C-

H), 1710, 1591 (C=C), 1506 (ArC-ArC), 1498 (N=N), 1425 (N=N), 1395 (Cl-ArC-ArC-Cl), 1359 (Cl-ArC-ArC-Cl), 1310 (Cl-ArC-ArC-Cl), 1256 (Cl-ArC-ArC-Cl), 1206 (Cl-ArC-ArC-Cl), 1154 (C-N), 1080 970, 814 (C-Cl), 731, 708. UV/Vis (THF): $\lambda_{max}$ (nm) ($\varepsilon$) 222 (52821), 388 (32775), 505 (6213), 693 (1909). EPR: g = 2.0026, $a(^1H)$ = 1.9 G, $\Delta H_{pp}$ = 1.1 G, $a(^{13}C_{Ar})$ = 12.6, 14.3 G, $a(^{13}C_\alpha)$ = 29.4 G.

**Synthesis of PTM-Dz and PTMH-Dz from the triazenes.** The respective triazenes were then converted to the diazonium salts by treatment with a strong acid following a previously reported protocol.[14]

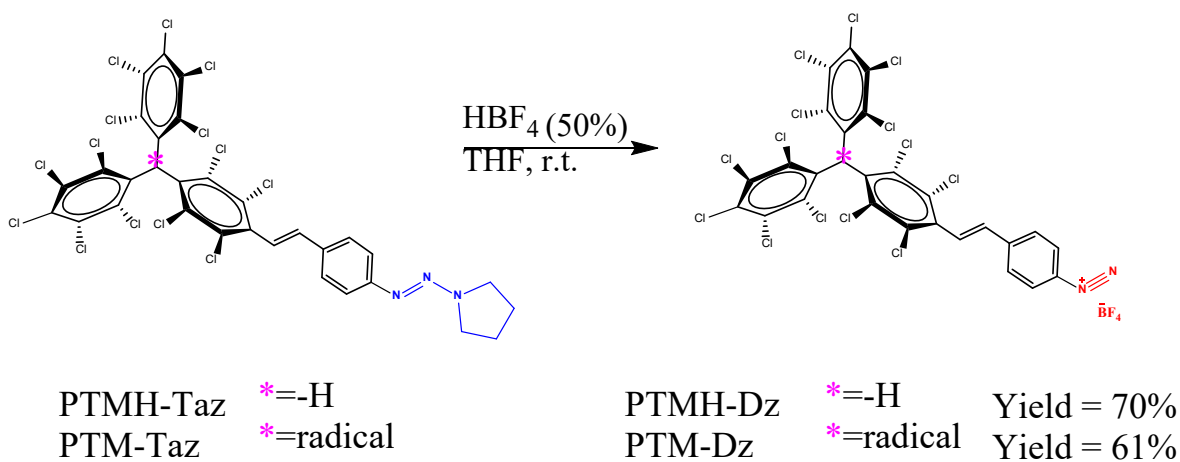

**Synthesis of PTM-Dz.** The synthesis was carried out under red-light illumination. **PTM-Taz** (64.8 mg, 0.07 mmol) was dissolved in dry THF (4 mL) and then an aqueous solution (1.0 mL) of tetrafluoroboric acid (50 wt-%) was added dropwise. The reaction mixture was stirred for about 20 min. The reaction was followed by UV/Vis spectroscopy. Then, the reaction crude was transferred to a 500 mL beaker and cold diethyl ether (200 mL) was added. The formation of orange powder suspended in the solution was observed. The suspension was maintained at -18ºC for 20 h until the orange powder precipitated. Excess solvent was decanted. The obtained powder was washed with cold diethyl ether (3 × 50 mL) and then dried under reduced pressure to yield the compound **PTM-Dz** as a red-orange powder (40 mg, 61 % yield). ATR-IR: $\nu$ (cm$^{-1}$) = 3104 (-Ar-H), 2256 (+N≡N), 1631(C=C), 1580 (C=C), 1508 (ArC-ArC), 1425, 1334 (Cl-ArC-ArC-Cl), 1318 (Cl-ArC-ArC-Cl), 1276 (Cl-ArC-ArC-Cl), 1260 (Cl-ArC-ArC-Cl), 1192 (Cl-ArC-ArC-Cl), 1158,

1074, 1050 (C-N), 1012, 970, 815 (C-Cl), 731, 709. UV/Vis (THF): λ$_{max}$ (nm) (ε) 228 (57162), 332 (21716), 386 (27740). CV (0.2 mol L$^{-1}$ TBAPF$_6$ in DCM, *vs*. Fc/Fc$^+$): E$_{1/2}$(radical/anion) = -0.45 V. EPR: g = 2.0026, *a*($^1$H) = 1.98 G, ΔH$_{pp}$ = 1.1 G, *a*($^{13}$C$_{Ar}$) = 12.6, 14.3 G, *a*($^{13}$C$_α$) = 29.4 G.

**Synthesis of PTMH-Dz.** The compound **PTMH-DZ** was synthesized from **PTMH-Taz** following the same procedure as described above. **PTMH-Dz** was obtained as a pale yellow powder in 70 % yield.

$^1$H NMR (600 MHz, Acetone-d$_6$): δ(ppm) = 8.94 (d, 2H, J=8.9 Hz, Ar-H), 8.39 (d, 2H, J=8.8 Hz, Ar-H), 7.77 (d, 1H, J=16.63 Hz, -C=C-H), 7.47 (d, 1H, J=16.66 Hz, -C=C-H), 7.12 (s, 1H alphaH). $^{13}$C NMR (151 MHz, Acetone-d$_6$): δ(ppm)= 149.73, 138.35, 137.55, 137.45, 136.47, 134.76, 134.75, 134.58, 134.44, 134.35, 133.35, 133.27, 130.53, 114.23(-ArC-N≡N+), 57.65 (alpha C). ATR-IR: ν (cm$^{-1}$) = 3105(ArC-H), 2256(+N≡N), 1633(C=C), 1578(C=C), 1532(ArC-ArC), 1425(Cl-ArC-ArC-Cl), 1359(Cl-ArC-ArC-Cl), 1322(Cl-ArC-ArC-Cl), 1298, 1241, 821, 808 (C-Cl), 725. UV/Vis (THF): λ$_{max}$ (nm) (ε) 226 (55512), 338 (19169). The raw data is provided on the following pages.

Since the **PTMH-Dz** has not been reported before, the recorded spectra are provided below.

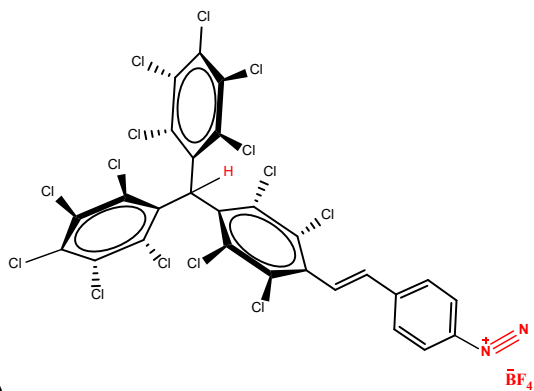

a)

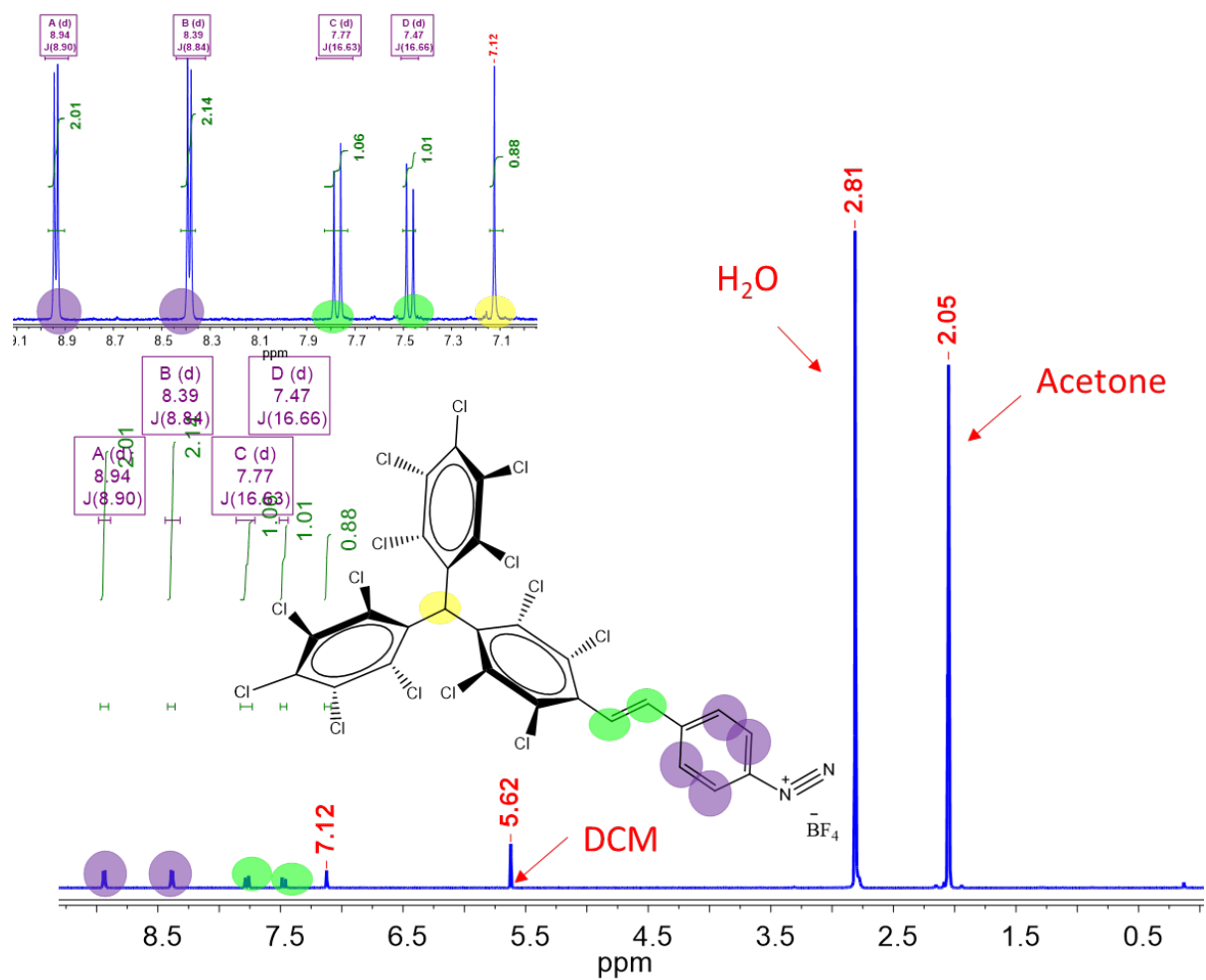

b)

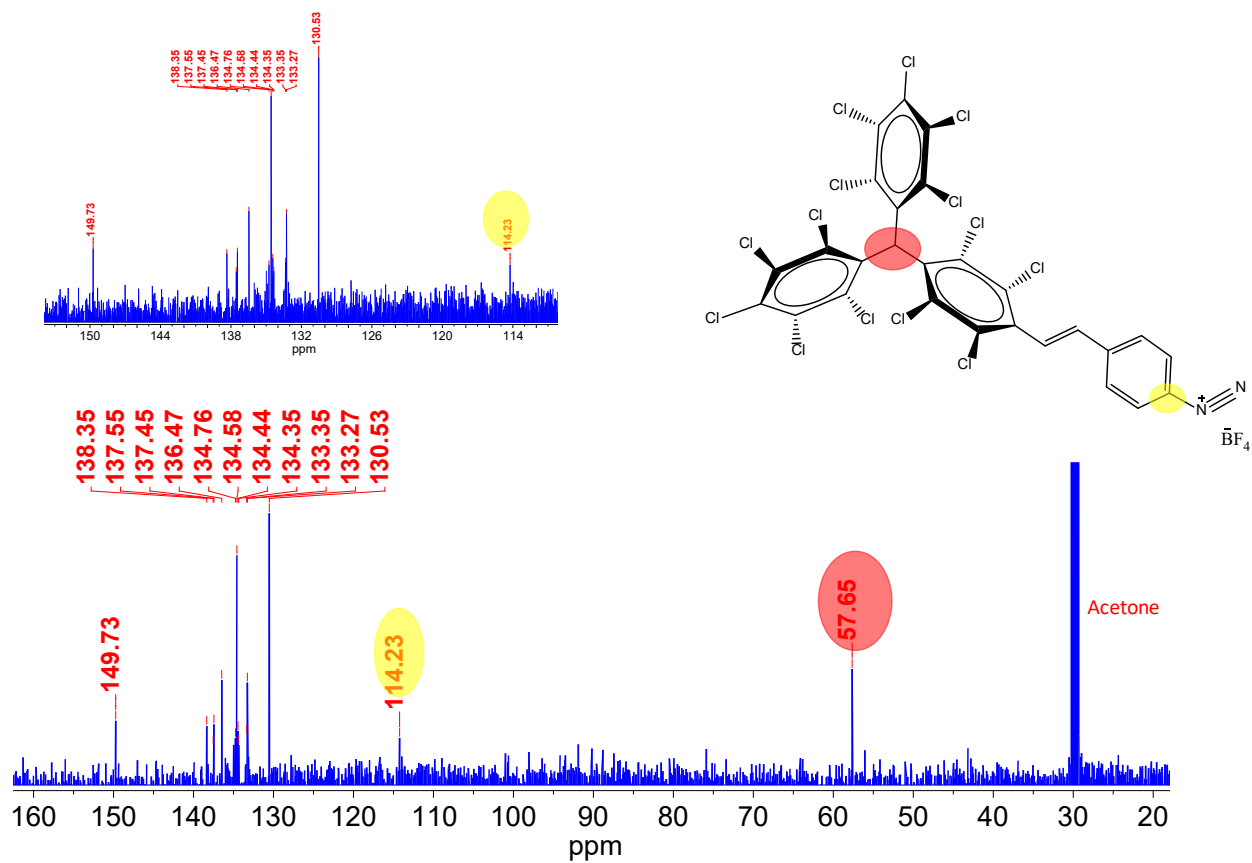

**Figure S10.** (a) $^1$H NMR spectrum of PTMH-Dz in acetone-$d_6$. Inset: zoom-in on the region of 7-9.1 ppm. (b) $^{13}$C NMR spectrum of PTMH-Dz in acetone-$d_6$. Inset: zoom-in on the region of 110-158 ppm.

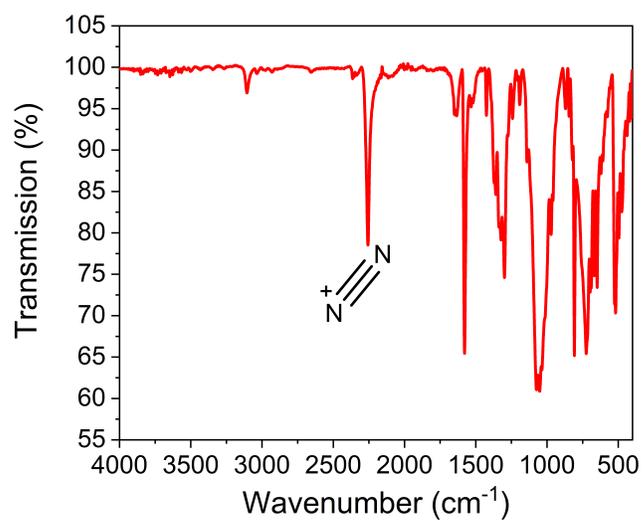

**Figure S11.** FTIR spectrum measured for the powder of **PTMH-Dz**.

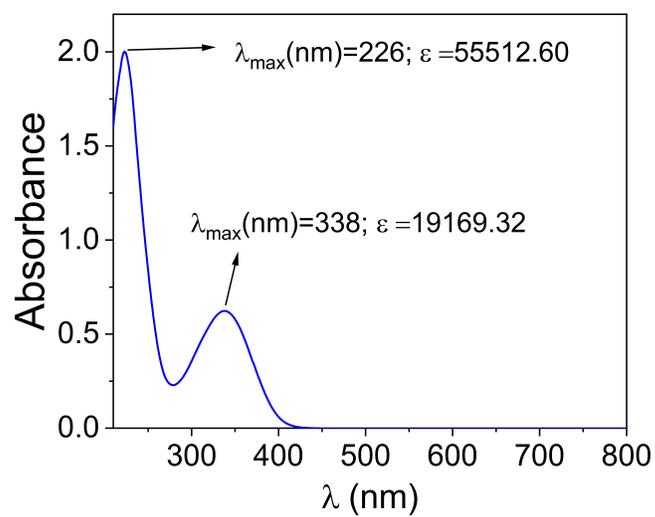

**Figure S12.** UV/vis absorption spectrum of **PTMH-Dz** in THF.

# REFERENCES

1. Berger, F. J.; Lüttgens, J.; Nowack, T.; Kutsch, T.; Lindenthal, S.; Kistner, L.; Müller, C. C.; Bongartz, L. M.; Lumsargis, V. A.; Zakharko, Y.; Zaumseil, J., Brightening of Long, Polymer-Wrapped Carbon Nanotubes by sp$^3$ Functionalization in Organic Solvents. *ACS Nano* **2019,** *13*, 9259-9269.
2. Fox, M. A.; Gaillard, E.; Chen, C.-C., Photochemistry of Stable Free Radicals: The Photolysis of Perchlorotriphenylmethyl Radicals. *J. Am. Chem. Soc.* **1987,** *109*, 7088-7094.
3. Ballester, M.; Castañer, J.; Riera, J.; Pujadas, J.; Armet, O.; Onrubia, C.; Rio, J. A., Inert Carbon Free Radicals. 5. Perchloro-9-phenylfluorenyl Radical Series. *J. Org. Chem.* **1984,** *49*, 770-778.
4. Lakowicz, J. R., *Principles of Fluorescence Spectroscopy*. Springer: New York, 2006.
5. Dexter, D. L., A Theory of Sensitized Luminescence in Solids. *J. Chem. Phys.* **1953,** *21*, 836-850.
6. Kwon, H.; Kim, M.; Nutz, M.; Hartmann, N. F.; Perrin, V.; Meany, B.; Hofmann, M. S.; Clark, C. W.; Htoon, H.; Doorn, S. K.; Högele, A.; Wang, Y., Probing Trions at Chemically Tailored Trapping Defects. *ACS Cent. Sci.* **2019,** *5*, 1786-1794.
7. Kim, Y.; Velizhanin, K. A.; He, X.; Sarpkaya, I.; Yomogida, Y.; Tanaka, T.; Kataura, H.; Doorn, S. K.; Htoon, H., Photoluminescence Intensity Fluctuations and Temperature-Dependent Decay Dynamics of Individual Carbon Nanotube sp$^3$ Defects. *J. Phys. Chem. Lett.* **2019,** *10*, 1423-1430.
8. Kim, M.; Adamska, L.; Hartmann, N. F.; Kwon, H.; Liu, J.; Velizhanin, K. A.; Piao, Y.; Powell, L. R.; Meany, B.; Doorn, S. K.; Tretiak, S.; Wang, Y., Fluorescent Carbon Nanotube Defects Manifest Substantial Vibrational Reorganization. *J. Phys. Chem. C* **2016,** *120*, 11268-11276.
9. Voisin, C.; Berger, S.; Berciaud, S.; Yan, H.; Lauret, J.-S.; Cassabois, G.; Roussignol, P.; Hone, J.; Heinz, T. F., Excitonic Signatures in the Optical Response of Single-Wall Carbon Nanotubes. *Phys. Status Solidi B* **2012,** *249*, 900-906.
10. Nutz, M.; Zhang, J.; Kim, M.; Kwon, H.; Wu, X.; Wang, Y.; Högele, A., Photon Correlation Spectroscopy of Luminescent Quantum Defects in Carbon Nanotubes. *Nano Lett.* **2019,** *19*, 7078-7084.
11. He, X.; Velizhanin, K. A.; Bullard, G.; Bai, Y.; Olivier, J. H.; Hartmann, N. F.; Gifford, B. J.; Kilina, S.; Tretiak, S.; Htoon, H.; Therien, M. J.; Doorn, S. K., Solvent- and Wavelength-Dependent Photoluminescence Relaxation Dynamics of Carbon Nanotube sp$^3$ Defect States. *ACS Nano* **2018,** *12*, 8060-8070.
12. Hartmann, N. F.; Velizhanin, K. A.; Haroz, E. H.; Kim, M.; Ma, X.; Wang, Y.; Htoon, H.; Doorn, S. K., Photoluminescence Dynamics of Aryl sp$^3$ Defect States in Single-Walled Carbon Nanotubes. *ACS Nano* **2016,** *10*, 8355-8365.
13. Uoyama, H.; Goushi, K.; Shizu, K.; Nomura, H.; Adachi, C., Highly Efficient Organic Light-Emitting Diodes from Delayed Fluorescence. *Nature* **2012,** *492*, 234-238.
14. Seber, G.; Rudnev, A. V.; Droghetti, A.; Rungger, I.; Veciana, J.; Mas-Torrent, M.; Rovira, C.; Crivillers, N., Covalent Modification of Highly Ordered Pyrolytic Graphite with a Stable Organic Free Radical by Using Diazonium Chemistry. *Chem. Eur. J.* **2017,** *23*, 1415-1421.